\documentclass[a4paper,11pt]{article}

\usepackage{amsmath}
\usepackage{amsfonts}
\usepackage{amssymb}
\usepackage{lscape}
\usepackage{rotating}

\begin{document}

\title{Relativistic phase shifts for Dirac particles    
interacting with weak gravitational fields in matter--wave  
interferometers\footnote{to appear in: C. L\"ammerzahl, C.W.F. Everitt, F.W. Hehl (eds.): {\it Gyros, Clocks, and Interferometers: Testing Relativistic Gravity in Space}, Springer--Verlag, Berlin 2000}}

\author{Christian J.~Bord\'{e}$^{1,2,3}$ 
and Jean--Claude Houard$^{1}$, 
and Alain Karasiewicz$^{1}$ \\
$^1$ Laboratoire de Gravitation et Cosmologie Relativistes, Universit\'e Pierre et Marie Curie, \\ 
4 Place Jussieu, 75252 Cedex 05, Paris, France \\ 
$^2$ Laboratoire de Physique des Lasers, UMR 7538 CNRS, Universit\'{e}  
Paris--Nord,\\
Avenue J.--B. Cl\'ement, 93430 Villetaneuse, France \\ 
$^3$ Institut f\"ur Quantenoptik, Universit\"at Hannover,
Welfengarten 1, \\ 30167 Hannover, Germany}

\maketitle

\begin{abstract}
We present a second--quantized field theory of massive spin one--half
particles or antiparticles in the presence of a weak gravitational field
treated as a spin two external field in a flat Minkowski background. We solve
the difficulties which arise from the derivative coupling and we are able to
introduce an interaction picture.
We derive expressions for the scattering amplitude and for the outgoing spinor
to first--order. In several appendices, the link with the canonical
approach in General Relativity is established and a generalized stationary phase
method is used to calculate the outgoing spinor. We show how our expressions
can be used to calculate and discuss phase shifts in the context of matter--wave interferometry (especially atom or antiatom interferometry).
In this way, many effects are introduced in a
unified relativistic framework, including spin--gravitation
terms: gravitational red shift, Thomas precession,
Sagnac effect, spin--rotation effect, orbital and spin Lense--Thirring effects,
de Sitter geodetic precession and finally the effect of gravitational waves.
A new analogy with the electromagnetic interaction is pointed out.
\end{abstract}

\section{Introduction}\index{spinhalfparticle@spin--$\frac{1}{2}$--particle}

The development of high accuracy \index{matter--wave interferometry}  atom interferometers, \index{atom interferometry} used as clocks in the
microwave or in the optical domain, as inertial sensors (gyros, gravimeters,
gradiometers...) or for the determination of atomic masses and of the fine
structure constant \cite
{Laurent,Borde1,Borde2,Borde4,Borde14,Borde15,Borde16,Borde17,Borde5,Borde6,Borde18},
requires now a framework to describe the interference of atom waves in a
rigorous way. On one hand, one needs to investigate general relativistic
effects including those involving the spin of the atoms and, on the other
hand, it is necessary to take into account the statistical properties
(bosonic or fermionic) of the interfering particles, given the development
of coherent atom wave sources and also for a proper treatment of the
detection noise. One would also like to be able to discuss the propagation
of antimatter in interferometers in the presence of gravitation and as
suggested in reference \cite{icols99} the properties of coherent antimatter
waves (generated by an antiatom laser such as an antihydrogen Bose--Einstein
condensate). This is possible only within relativistic quantum field theory.
Atoms (or antiatoms), in a given eigenstate of the internal atomic
Hamiltonian, are considered as elementary particles having a rest mass fixed
by the energy of the atomic level and a spin equal to the total angular
momentum of the atom in that level. In this paper, we shall consider only
Dirac particles for illustration. This is the simplest example of
particles with spin which still contains most interesting effects related to
spin and applies to neutron or electron interferometry as a special case.
The generalization to others spin values is possible along similar lines
with Dirac--type equations (Bargmann--Wigner \cite{BW}, de Broglie fusion
method \cite{fusion}, Durand \cite{Durand}). Our overall goal is to
introduce gravitation and general relativistic effects at the quantum level
of modern atomic physics and quantum optics experiments. For this, we
propose an extension of our first paper on atom interferometry in General
Relativity \cite{BKT}, which includes now a second--quantization scheme for
the atom waves in the presence of gravitational and electromagnetic fields.
The point of view adopted in this paper is the extrinsic point of view using
purely quantum field theory in a flat Minkowski background. The connection
with the canonical intrinsic approach, using Dirac equation in curved
space--time, is made in Appendices A and B. The reader who wishes to start
with this canonical approach is thus invited to read first these appendices.

In the main text, we begin right away with the minimal coupling Lagrangian
in flat space--time, which is identical to the one derived in curved
space--time for a standard choice of tetrads. Then, we proceed with the
quantization of the Dirac field and we emphasize the difficulties which
arise because of the derivative coupling. These difficulties are solved in a
consistent scheme which allows also to define an interaction picture. The
evolution operator and the $S$--matrix are constructed and we demonstrate
explicitly a conjecture of Gupta. These results are used to derive formulas
for the transition amplitude and for the outgoing spinor in the weak--field
approximation, first in configuration space and second in the momentum
representation. An expansion in the perturbation wave vector $(\hbar k/mc)$
is used to retrieve various physical effects, some of which are well--known.
A new analogy with the electromagnetic interaction is presented which
includes all components of the field $h^{\mu \nu }$ and generalizes
gravitoelectric and gravitomagnetic interactions. In Appendices C and D,
calculations of the outgoing spinor are sketched, first with a generalized
stationary phase method in configuration space and second in the momentum
representation.

\section{Lagrangian theory}

Considered as a field theory in flat spacetime, the theory describing the
interaction of matter with a given gravitational field will be defined by a
Lagrangian density of the following form \cite{G1,F}: 
\begin{equation}
\mathcal{L}=\mathcal{L}_{0}-{\textstyle\frac{1}{2}}h^{\mu \nu }T_{\mu \nu
}\,,  \label{lag}
\end{equation}
where $h^{\mu \nu }$ is the given external field, and where $\mathcal{L}_{0}$
is the free Lagrangian density of the matter field and $T_{\mu \nu }$ the
corresponding stress--energy tensor. For a Dirac field, one has respectively%
\footnote{%
The conventions used here for the metric and for the Dirac equation \index{Dirac equation} and
matrices are generally those of \cite{BD}. In particular, the signature of
the metric is taken as $(+,-,-,-)$. Greek indices $\mu ,\nu ,...$ run from 0
to 3 and latin indices run from 1 to 3. The space--time 4--vector is written $%
x=(x^{0},\vec{x})=(ct,\vec{x})$.\ \ The partial derivatives with the right
arrow act on the right and those with the left arrow on the left.}, in
symmetrical form, 
\begin{align}
\mathcal{L}_{0}& =\frac{\hbar c}{2}\left[ \overline{\Psi }\left( i\gamma
^{\mu }\overset{\rightarrow }{\partial _{\mu }}-\frac{mc}{\hbar }\right)
\Psi +\overline{\Psi }\left( -i\gamma ^{\mu }\overset{\leftarrow }{\partial
_{\mu }}-\frac{mc}{\hbar }\right) \Psi \right] \,,  \label{lag0} \\
T_{\mu \nu }& =-\eta _{\mu \nu }\mathcal{L}_{0}+\frac{\hbar c}{4}\left[ 
\overline{\Psi }\left( i\gamma _{\mu }\overrightarrow{\partial _{\nu }}-i%
\overleftarrow{\partial _{\nu }}\gamma _{\mu }\right) \Psi +\overline{\Psi }%
\left( i\gamma _{\nu }\overrightarrow{\partial _{\mu }}-i\overleftarrow{%
\partial _{\mu }}\gamma _{\nu }\right) \Psi \right] \,.  \label{T}
\end{align}
The total Lagrangian density (\ref{lag}) then becomes 
\begin{equation}
\mathcal{L}=\left( 1+{\textstyle\frac{1}{2}}h\right) \mathcal{L}_{0}-{%
\textstyle\frac{1}{4}}i\hbar ch^{\mu \nu }\,\overline{\Psi }\left( \gamma
_{\mu }\overrightarrow{\partial }_{\nu }-\overleftarrow{\partial }_{\nu
}\gamma _{\mu }\right) \Psi ,  \label{Lag}
\end{equation}
where\footnote{The term $(h/2)\mathcal{L}_{0}$ in (\ref{Lag}) comes from the term
containing $\mathcal{L}_{0}$ in (\ref{T}), and was omitted in \cite{G2}.
However, $\mathcal{L}_{0}$ which vanishes when the free Dirac equation \index{Dirac equation} is
satisfied, does not vanish here.} $h=h^{\mu }{}_{\mu }=\eta _{\mu \nu
}h^{\mu \nu }$. This Lagrangian density can also be considered as obtained
from the Lagrangian density valid in General Relativity for the interaction
of the Dirac field with a prescribed gravitational field in the linear
approximation (see the Appendices).

The equations for $\Psi $ and $\overline{\Psi }$ derived from (\ref{Lag})
are the following\footnote{%
If one considers these equations as first--order equations with respect to
the $h^{\mu \nu }$'s, the factor $(1+h/2)$ in the first term can be replaced
by 1. But it can be shown that the equations so obtained cannot be derived
from a Lagrangian, if this latter is restricted to depend linearly on $\Psi $
and $\overline{\Psi }$, and to admit first--order derivatives only.}: 
\begin{align}
\lbrack \left( 1+{\textstyle\frac{1}{2}}h\right) (i\gamma ^{\mu }%
\overrightarrow{\partial }_{\mu }-{\textstyle mc/\hbar })-\frac{i}{2}h^{\mu
\nu }\gamma _{\mu }\overrightarrow{\partial }_{\nu }-\frac{i}{4}\partial
_{\nu }h^{\mu \nu }\gamma _{\mu }+\frac{i}{4}\partial _{\mu }h\gamma ^{\mu
}]\Psi & =0,  \label{eqpsi} \\
\overline{\Psi }[(-i\gamma ^{\mu }\overleftarrow{\partial }_{\mu }-{%
\textstyle}mc/\hbar )\left( 1+{\textstyle\frac{1}{2}}h\right) +\frac{i}{2}%
\overleftarrow{\partial }_{\nu }\gamma _{\mu }h^{\mu \nu }+\frac{i}{4}%
\partial _{\nu }h^{\mu \nu }\gamma _{\mu }-\frac{i}{4}\partial _{\mu
}h\gamma ^{\mu }]& =0.  \label{eqpsibar}
\end{align}
They admit the conserved current 
\begin{equation}
j^{\mu }=c\overline{\Psi }[\gamma ^{\mu }+{\textstyle\frac{1}{2}}h\gamma
^{\mu }-{\textstyle\frac{1}{2}}h^{\mu \nu }\gamma _{\nu }]\Psi ,  \label{j}
\end{equation}
which coincides with the usual Dirac current when the gravitational field
vanishes.

In order to stress some other differences and analogies with
electromagnetism, we may write the equation for $\Psi $ with a covariant
derivative in the usual sense of non--Abelian gauge field theories in flat
space--time

\begin{equation}
i\gamma ^{\nu }(\overrightarrow{\partial }_{\nu }+\frac{i}{4}\sigma
^{\lambda \mu }\partial _{\lambda }h_{\mu \nu }-\frac{1}{2}h_{\nu
}{}^{\alpha }\overrightarrow{\partial }_{\alpha })\Psi -\frac{mc}{{\hbar }}%
\Psi =0  \label{covariantderivative}
\end{equation}
where 
\begin{equation}
\sigma ^{\mu \nu }=\frac{i}{2}\left( \gamma ^{\mu }\gamma ^{\nu }-\gamma
^{\nu }\gamma ^{\mu }\right) \,
\end{equation}
and where the factor $\left( 1+{\textstyle\frac{1}{2}}h\right) $ has been
removed. The Poincar\'{e} generators are associated with gauge fields and a
local gauge invariance\footnote{%
Differences with usual Yang--Mills theories come from the non--commutation of
the Lorentz generators with the Dirac matrices and from the fact that the
translation generators act on space--time itself.}

\begin{eqnarray}
\Psi ^{\prime }(x) &=&\left[ 1+\frac{i}{8}\sigma ^{\lambda \mu }\left(
\partial _{\lambda }\xi _{\mu }-\partial _{\mu }\xi _{\lambda }\right) -\xi
^{\lambda }\partial _{\lambda }\right] \Psi (x) \\
h_{\mu \nu }^{\prime }(x) &=&h_{\mu \nu }(x)-\partial _{\mu }\xi _{\nu
}-\partial _{\nu }\xi _{\mu }\quad .
\end{eqnarray}

\section{Quantization}

We want now to proceed with the quantization of the field $\Psi $ submitted
to the interaction defined by (\ref{Lag}). When they are applied directly,
the standard methods lead to some difficulties coming from the presence of a
derivative coupling in the Lagrangian. These problems are, first, briefly
discussed, then, a solution is presented allowing the quantization together
with the definition of the interaction picture.

\subsection{Difficulties with the derivative coupling}

Two methods can be used, \emph{a priori}, according to whether the
Lagrangian is taken under a symmetrical or a asymmetrical form.

\subsubsection{Symmetrical Lagrangian.}

Starting from the symmetrical Lagrangian (\ref{Lag}), the usual
anticommutation relations will be obtained from the following expression of
the conjugate momentum 
\begin{equation}
\Pi =2\,\frac{\partial \mathcal{L}}{\partial \dot{\Psi}}\, , \qquad\text{
with} \qquad \dot{\Psi}=\partial _{t}\Psi \, ,  \label{pi}
\end{equation}
which gives 
\begin{equation}
\Pi =i{\hbar }\overline{\Psi }[(1+ {\textstyle \frac{1}{2}} h)\gamma ^{0}- {\textstyle \frac{1}{2}} h^{0\mu} \gamma_\mu],
\end{equation}
and 
\begin{equation}
\{\Psi _{\alpha }(x),\Pi _{\beta }(y)\}_{x^{0}=y^{0}}=i\hbar \delta _{\alpha
\beta }\,\delta (\vec{x}-\vec{y})\,
\end{equation}
where $\alpha $ and $\beta $ are spinorial indices.

By introducing the matrix (depending on the coordinates) 
\begin{equation}
\gamma ^{-1}=(1+ {\textstyle \frac{1}{2}} h)\gamma ^{0}- {\textstyle \frac{1}{2}} h^{0\mu }\gamma_\mu \, ,
\end{equation}
these formulas can be rewritten 
\begin{equation}
\Pi =i{\hbar }\overline{\Psi }\gamma ^{-1},
\end{equation}
and 
\begin{equation}
\{\Psi _{\alpha }(x),\overline{\Psi }_{\beta }(y)\}_{x^{0}=y^{0}}=(\gamma
(x))_{\alpha \beta }\,\delta (\vec{x}-\vec{y}).  \label{RC}
\end{equation}
For the free field the matrix $\gamma (x)$ reduces to $\gamma ^{0}$.
However, it appears that the Heisenberg equations, which are generally
equivalent to the equations of motion, are not satisfied here under their
most usual forms. To write the formulas in a condensed form, let us
introduce the two operators 
\begin{align}
\overrightarrow{\mathcal{D}}& =\hbar c\left[ (1+ {\textstyle \frac{1}{2}} h)(-i\gamma^k \overrightarrow{\partial _{k}}+mc/\hbar )+\frac{i}{2}h^{\mu k}\gamma_\mu \overrightarrow{\partial _{k}}+\frac{i}{4}(\partial _{\nu }h^{\mu \nu}-\partial ^{\mu }h)\gamma _{\mu }\right] \,, \\
\overleftarrow{\mathcal{D}}& =\hbar c\left[ (i\gamma ^{k}\overleftarrow{\partial_{k}}+mc/\hbar )(1+ {\textstyle \frac{1}{2}} h) - \frac{i}{2}\overleftarrow{\partial _{k}} \gamma_{\mu} h^{\mu k}-\frac{i}{4}(\partial _{\nu }h^{\mu \nu }-\partial
^{\mu} h)\gamma _{\mu }\right] \,.
\end{align}
One then finds that the Hamiltonian density $\Pi \dot{\Psi}-\mathcal{L}$ can
be written as 
\begin{equation}
\mathcal{H}=\frac{1}{2}[\overline{\Psi }\overrightarrow{\mathcal{D}}\Psi +%
\overline{\Psi }\overleftarrow{\mathcal{D}}\Psi ],
\end{equation}
while the field equation reads 
\begin{equation}
i\hbar c\gamma ^{-1}\partial _{0}\Psi =\overrightarrow{\mathcal{D}}\Psi .
\end{equation}
The relation $\overline{\overrightarrow{\mathcal{D}}\Psi }=\overline{\Psi }%
\overleftarrow{\mathcal{D}}$ and the formula $(\gamma ^{-1})^{\dag }=\gamma
^{0}\gamma ^{-1}\gamma ^{0}$ then give the conjugate equation 
\begin{equation}
-i\hbar c\partial _{0}\overline{\Psi }\,\gamma ^{-1}=\overline{\Psi }%
\overleftarrow{\mathcal{D}}.
\end{equation}
Using an integration by parts, the total Hamiltonian $H$ can be transformed
into 
\begin{equation}
H=\int (d^{3}x)\overline{\Psi }\overrightarrow{\mathcal{D}}\Psi +\frac{%
i\hbar c}{2}\int (d^{3}x)\overline{\Psi }(\partial _{0}\gamma ^{-1})\Psi .
\label{hpsi}
\end{equation}
The anticommutation relation (\ref{RC}) then gives the commutator 
\begin{equation}
\left[ H,\Psi \right] =-\gamma \overrightarrow{\mathcal{D}}\Psi -\frac{i}{2}%
\hbar c(\gamma \partial _{0}\gamma ^{-1})\Psi \,,
\end{equation}
or, given the field equation, 
\begin{equation}
\left[ H,\Psi \right] =-i\hbar c\partial _{0}\Psi -\frac{i}{2}\hbar c(\gamma
\partial _{0}\gamma ^{-1})\Psi .  \label{HPsi}
\end{equation}
Similarly, one has 
\begin{equation}
\left[ H,\overline{\Psi }\right] =-i\hbar c\partial _{0}\overline{\Psi }-%
\frac{i}{2}\hbar c\overline{\Psi }(\partial _{0}\gamma ^{-1})\gamma ,
\label{HPsibar}
\end{equation}
then 
\begin{equation}
\left[ H,\Pi \right] =-i\hbar c\partial _{0}\Pi +\frac{i}{2}\hbar c\Pi
\gamma \partial _{0}\gamma ^{-1}.  \label{HPi}
\end{equation}
The second terms in the right members of (\ref{HPsi}), (\ref{HPsibar}) and (%
\ref{HPi}) are unusual, since the fields $\Psi ,\;\overline{\Psi }$ and $\Pi 
$ behave as if, considered as functions of some fundamental dynamical
variables, they were also explicitly dependent on the time. From the
expression of $\Pi $, it is in fact obvious that, among $\overline{\Psi }$
and $\Pi $ one of them, at least, explicitly depends on time. But this is
not obvious for~$\Psi $.

\subsubsection{Asymmetrical Lagrangian.}

The preceding difficulty takes another form if, as it is more usual \cite
{BD2}, the Lagrangian (\ref{Lag}) is replaced by the asymmetrical one 
\begin{align}
\mathcal{L^{\prime }}& = \mathcal{L}+\frac{i\hbar }{2}\partial _{\mu }j^{\mu }
\notag \\
& = \hbar c\, \overline{\Psi }\left[ (1+{\textstyle\frac{1}{2}}h)(i\gamma ^{\mu } \overrightarrow{\partial_\mu} - \!{\textstyle mc/\hbar}) \Psi -\frac{i}{2} h^{\mu \nu }\gamma _{\mu }\overrightarrow{\partial _{\nu }}\Psi -\frac{i}{4} \partial _{\nu }h^{\mu \nu }\gamma _{\mu }\Psi + \!\frac{i}{4}\partial _{\mu
}h\gamma ^{\mu }\Psi\right]\!.
\end{align}
The field equations are left unchanged, and the conjugate field is again: 
\begin{equation}
\Pi ^{\prime }=\frac{\partial \mathcal{L^{\prime }}}{\partial \dot{\Psi}}=i{%
\hbar }\overline{\Psi }\gamma ^{-1}.  \label{piprime}
\end{equation}
Since $\Pi ^{\prime }=\Pi $, the anticommutation relation of $\Psi $ and $%
\overline{\Psi }$ is identical to (\ref{RC}), but the Hamiltonian is now 
\begin{equation}
H^{\prime }=\int (d^{3}x)\overline{\Psi }\overrightarrow{\mathcal{D}}\Psi
=\int (d^{3}x)\overline{\Psi }\overleftarrow{\mathcal{D}}\Psi -i\hbar c\int
(d^{3}x)\overline{\Psi }(\partial _{0}\gamma ^{-1})\Psi .
\end{equation}
It follows that the usual Heisenberg equations are satisfied. In fact, one
has 
\begin{align}
\left[ H^{\prime },\Psi \right] & =-\gamma \overrightarrow{\mathcal{D}}\Psi
=-i\hbar c\partial _{0}\Psi \,, \\
\left[ H^{\prime },\overline{\Psi }\right] & =\overline{\Psi }\overleftarrow{%
\mathcal{D}}\gamma -i\hbar c\overline{\Psi }(\partial _{0}\gamma
^{-1})\gamma =-i\hbar c\partial _{0}\overline{\Psi }-i\hbar c\overline{\Psi }%
(\partial _{0}\gamma ^{-1})\gamma ,  \label{HprimePsibar}
\end{align}
then 
\begin{equation}
\left[ H^{\prime },\Pi ^{\prime }\right] =-i\hbar c\partial _{0}\Pi ^{\prime
}.
\end{equation}
The equations satisfied by $\Psi $ and $\Pi ^{\prime }$ are those of
variables having no explicit time dependence, which is the rule for
canonical variables. On the contrary, expression (\ref{piprime}) shows that $%
\overline{\Psi }$ depends explicitly on time since $\gamma ^{-1}$ does, and
equation (\ref{HprimePsibar}) is in agreement with this dependence.

However, the Hamiltonian $H^{\prime }$ is not Hermitian. In fact, since one
has $(\overline{\Psi }\overrightarrow{\mathcal{D}}\Psi )^{\dag }=\overline{%
\Psi }\overleftarrow{\mathcal{D}}\Psi $, it follows that 
\begin{equation}
H^{\prime }{}^{\dag }=\int (d^{3}x)\overline{\Psi }\overleftarrow{\mathcal{D}%
}\Psi =H^{\prime }+i\hbar c\int (d^{3}x)\overline{\Psi }(\partial _{0}\gamma
^{-1})\Psi .
\end{equation}
The asymmetry of the equations for $\Psi $ and $\overline{\Psi }$ is a
consequence of this lack of Hermiticity.

\subsubsection{Trouble with the interaction picture.}

A common shortcoming of the two preceding methods of quantization is the
absence of a coherent definition of the interaction picture. In fact, if
this picture was defined, there would be a unitary operator $U$, such that
the corresponding field variables $\psi $ and $\varpi $ would be given by 
\begin{equation}
\psi =U\Psi U^{-1}\,,\qquad \varpi =U\Pi U^{-1}.  \label{U}
\end{equation}
Moreover, these variables would be free fields so that one would have the
relation $\varpi =i\hbar \overline{\psi }\gamma _{0}$. Such a relation is
not compatible with (\ref{U}) since one has $\Pi =i\hbar \overline{\Psi }%
\gamma ^{-1}$, %ligne 301
$\gamma ^{-1}\neq \gamma _{0}$, and since $U$ must be unitary.

In what follows, the quantization is defined in a way such that the
preceding difficulties do not appear. In particular, the interaction picture
will be defined, allowing the construction of the transition probabilities
and that of the S matrix.

\subsection{A coherent method of quantization}

The afore--mentioned problems will be solved by a change of variables
eliminating the derivative coupling. The Lagrangian density (\ref{Lag}) can
be written under the form 
\begin{equation}
\mathcal{L}=\frac{i\hbar c}{2}\left( \overline{\Psi }\gamma^{-1}(\partial
_{0}\Psi )-(\partial _{0}\overline{\Psi })\gamma ^{-1}\Psi \right) -\frac{1}{%
2}\overline{\Psi }(\overrightarrow{\mathcal{D}}+\overleftarrow{\mathcal{D}}%
)\Psi .  \label{lagg}
\end{equation}
%ligne 321
Let us introduce the new field $\Theta $ by the formula $\Psi =\Lambda
\Theta $, where $\Lambda $ is a matrix to\ be determined, which depends on
the coordinates. One has 
\begin{equation}
\overline{\Psi }\gamma ^{-1}(\partial _{0}\Psi )=\overline{\Theta }\gamma
^{0}\Lambda ^{\dag }\gamma ^{0}\gamma ^{-1}\Lambda (\partial _{0}\Theta )+%
\overline{\Theta }\gamma ^{0}\Lambda ^{\dag }\gamma ^{0}\gamma
^{-1}(\partial _{0}\Lambda )\Theta .
\end{equation}
The terms containing the time derivatives of $\Theta $ and $\overline{\Theta 
}$ in (\ref{lagg}) will be those of the free Dirac Lagrangian if one has $%
\gamma ^{0}\Lambda ^{\dag }\gamma ^{0}\gamma ^{-1}\Lambda =\gamma ^{0}$ or,
assuming the inversibility of $\Lambda $, if $\gamma ^{0}\gamma
^{-1}=\Lambda ^{\dag -1}\Lambda ^{-1}$. By writing the matrix $\Lambda $
under the form $\Lambda =MU$ where $M$ is Hermitian and positive \cite{Bron}
and $U$ unitary, the preceding equation becomes 
\begin{equation}
\gamma ^{0}\gamma ^{-1}=(M^{-1})^{2}= I + {\textstyle \frac{1}{2}} h - {\textstyle \frac{1}{2}} h^{0\mu
}\gamma ^{0}\gamma _{\mu }.  \label{M}
\end{equation}
If $h_{\mu \nu }$ is sufficiently small, this matrix is inversible and
defines a unique matrix $M$ positive--definite \cite{Bron}, while $U$ may be
arbitrary. However, to ensure that the asymptotic states deduced from $%
\Theta $ or $\Psi $ have the same physical interpretation, it is necessary
that $U$ goes to the identity when $t\rightarrow \pm \infty $. More
precisely, we will impose the condition $U=I$, by which, when $h_{\mu \nu
}=0 $, the field $\Theta $ is a free field as $\Psi $.

The Lagrangian density takes the following form as a function of $\Theta $: 
\begin{equation}
\mathcal{L}=\hbar c\left[ \frac{i}{2}\left( \overline{\Theta }\gamma
^{0}(\partial _{0}\Theta )-(\partial _{0}\overline{\Theta })\gamma
^{0}\Theta \right) +\frac{i}{2}\left( \overline{\Theta }\Gamma ^{k}(\partial
_{k}\Theta )-(\partial _{k}\overline{\Theta })\Gamma ^{k}\Theta \right) +%
\frac{1}{2}\overline{\Theta }\Gamma \Theta \right] ,  \label{lagtheta}
\end{equation}
where $\Gamma ^{k}$ and $\Gamma $ are defined by 
\begin{align}
\Gamma ^{k}& =\gamma ^{0}\Lambda ^{\dag }\gamma ^{0}\left[ (1+{\textstyle%
\frac{1}{2}}h)\gamma ^{k}-{\textstyle\frac{1}{2}}h^{\mu k}\gamma _{\mu }%
\right] \Lambda ,  \label{gamk} \\
\Gamma & =i\left( \gamma ^{0}\Lambda ^{-1}\partial _{0}\Lambda +\Gamma
^{k}\Lambda ^{-1}\partial _{k}\Lambda \right)  \notag \\
& \quad -i\gamma ^{0}\left( (\partial _{0}\Lambda ^{\dag })\Lambda ^{\dag
-1}\gamma ^{0}+(\partial _{k}\Lambda ^{\dag })\Lambda ^{\dag -1}(\gamma
^{0}\Gamma ^{k}\gamma ^{0})\right) \gamma ^{0}-2\frac{mc}{\hbar }(1+{%
\textstyle\frac{1}{2}}h)\gamma ^{0}\Lambda ^{\dag }\gamma ^{0}\Lambda \,.
\label{B_gam}
\end{align}
Let us note the Hermiticity relations 
\begin{equation}
(\Gamma ^{k})^{\dag }=\gamma ^{0}\Gamma ^{k}\gamma ^{0}\,,\qquad \Gamma
^{\dag }=\gamma ^{0}\Gamma \gamma ^{0}\,.
\end{equation}
The field equations, equivalent to those of $\Psi $ and $\overline{\Psi }$,
are now 
\begin{align}
i\left( \gamma ^{0}(\partial _{0}\Theta )+\Gamma ^{k}(\partial _{k}\Theta
)\right) +\frac{1}{2}\left( \Gamma +i(\partial _{k}\Gamma ^{k})\right)
\Theta & = 0\,,  \label{eqtheta} \\
-i\left( (\partial_0 \overline{\Theta })\gamma ^{0}+(\partial _{k}%
\overline{\Theta })\Gamma ^{k}\right) +\frac{1}{2}\overline{\Theta }\left(
\Gamma -i(\partial _{k}\Gamma ^{k})\right) & =0,  \label{eqthetabar}
\end{align}
while the current reads 
\begin{equation}
j^{\mu} = c \, \overline{\Theta} \gamma^0 \Lambda^{\dag} \gamma^0 \left[(1+ {\textstyle \frac{1}{2}} h) \gamma^\mu - {\textstyle \frac{1}{2}} h^{\mu\nu}\gamma_\nu\right]
\Lambda \Theta \,,
\end{equation}
or, more explicitly\footnote{%
The expression of $j^{0}$ is identical to that of the free--field case. In
Appendix B this property is taken as a condition allowing the introduction
of the field $\Theta $ in the framework of the linearized theory of General
Relativity.}, 
\begin{equation}
j^{0}=c\overline{\Theta }\gamma ^{0}\Theta ,\hspace{20mm}j^{k}=c\overline{%
\Theta }\Gamma ^{k}\Theta \,.
\end{equation}

The conjugate momentum $\Pi _{\Theta }$ of $\Theta $ has the same form as in
the free--field case 
\begin{equation}
\Pi _{\Theta }=2\,\frac{\partial \mathcal{L}}{\partial \dot{\Theta}}=i\hbar 
\overline{\Theta }\gamma ^{0}\,.
\end{equation}
It follows that the anticommutation relation of $\Theta $ and $\overline{%
\Theta }$ is the usual one 
\begin{equation}
\{\Theta _{\alpha }(x),\overline{\Theta }_{\beta
}(y)\}_{x^{0}=y^{0}}=\,\gamma _{\alpha \beta }^{0}\,\delta (\vec{x}-\vec{y}).
\end{equation}
It is equivalent to the anticommutation relation (\ref{RC}) of $\Psi $ and $%
\overline{\Psi }$. The Hamiltonian density $\mathcal{H}_{\Theta }$ now reads 
\begin{equation}
\mathcal{H}_{\Theta }=-\frac{i\hbar c}{2}\left( \overline{\Theta }\Gamma
^{k}(\partial _{k}\Theta )-(\partial _{k}\overline{\Theta })\Gamma
^{k}\Theta \right) -\frac{\hbar c}{2}\overline{\Theta }\Gamma \Theta ,
\end{equation}
which gives for the total Hamiltonian 
\begin{equation}
H_{\Theta }=-\frac{\hbar c}{2}\int (d^{3}x)\left[ i\left( \overline{\Theta }%
\Gamma ^{k}(\partial _{k}\Theta )-(\partial _{k}\overline{\Theta })\Gamma
^{k}\Theta \right) +\overline{\Theta }\Gamma \Theta \right] ,  \label{Htot}
\end{equation}
or, equivalently, 
\begin{align}
H_{\Theta }& =-\hbar c\int (d^{3}x)\overline{\Theta }\left[ i\Gamma
^{k}(\partial _{k}\Theta )+\frac{1}{2}\left( \Gamma +i(\partial _{k}\Gamma
^{k})\right) \Theta \right] ,  \label{htheta} \\
& =\hbar c\int (d^{3}x)\left[ i(\partial _{k}\overline{\Theta })\Gamma ^{k}-%
\frac{1}{2}\overline{\Theta }\left( \Gamma -i(\partial _{k}\Gamma
^{k})\right) \right] \Theta .
\end{align}
This operator is Hermitian, and from the preceding expressions and, from the
field equations (\ref{eqtheta}) and (\ref{eqthetabar}), one checks the
Heisenberg equations 
\begin{equation}
\left[ H_{\Theta },\Theta \right] =-i\hbar c\partial _{0}\Theta \,,\qquad %
\left[ H_{\Theta },\overline{\Theta }\right] =-i\hbar c\partial _{0}%
\overline{\Theta }\,,\qquad \left[ H_{\Theta },\Pi _{\Theta }\right]
=-i\hbar c\partial _{0}\Pi _{\Theta }\,.
\end{equation}
The difficulties discussed in Section 3.1 have disappeared. In particular,
the basic variables $\Theta $, $\overline{\Theta }$ and $\Pi _{\Theta }$
look like variables having no explicit dependence on time. On the contrary,
the variables $\Psi $, $\overline{\Psi }$ and $\Pi $, initially considered,
depend explicitly on time, since one has 
\begin{equation}
\Psi =\Lambda \Theta \,,\qquad \overline{\Psi }=\overline{\Theta }\gamma
^{0}\Lambda ^{\dag }\gamma ^{0}\,,\qquad \Pi =\Pi _{\Theta }\Lambda ^{-1},
\end{equation}
and since the matrix $\Lambda $ generally depends on time.

Let us remark, however, that the Hamiltonian $H_{\Theta }$ differs from the
Hamiltonian $H$ introduced at the beginning in terms of $\Psi $. It is
convenient to consider $H$ from (\ref{hpsi}) as the integral of the density 
\begin{equation}
\mathcal{H}_{\Psi }=\overline{\Psi }\left[ \overrightarrow{\mathcal{D}}+%
\frac{i\hbar c}{2}(\partial _{0}\gamma ^{-1})\right] \Psi ,
\end{equation}
and $H_{\Theta }$ from (\ref{htheta}) as the integral of the density 
\begin{equation}
\mathcal{H}_{\Theta }=\hbar c\overline{\Theta }\left[ -i\Gamma ^{k}%
\overrightarrow{\partial _{k}}-\frac{1}{2}\left( \Gamma +i(\partial
_{k}\Gamma ^{k})\right) \right] \Theta .
\end{equation}
A rather tedious transformation of the former expression, using the
expressions of $\gamma ^{-1}$, $\Gamma ^{k}$, $\overrightarrow{\mathcal{D}}$%
, and the definition of $\Theta $, leads to the formula 
\begin{equation}
\mathcal{H}_{\Psi }=\mathcal{H}_{\Theta }+\frac{i\hbar c}{2}\overline{\Theta 
}\gamma ^{0}\left[ \Lambda ^{-1}(\partial _{0}\Lambda )-(\partial
_{0}\Lambda ^{\dag })\Lambda ^{\dag -1}\right] \Theta .
\end{equation}
Conversely, one has 
\begin{equation}
\mathcal{H}_{\Theta }=\mathcal{H}_{\Psi }+\frac{i\hbar c}{2}\overline{\Psi }%
\gamma ^{0}\left[ \Lambda ^{\dag -1}(\partial _{0}\Lambda ^{-1})-(\partial
_{0}\Lambda ^{\dag -1})\Lambda ^{-1}\right] \Psi \,.
\end{equation}
With the choice made above of the matrix $U$, one has $\Lambda =M=M^{\dag }$%
, with $M$ defined by (\ref{M}). The two preceding formulas can then be
written 
\begin{align}
\mathcal{H}_{\Psi }& =\mathcal{H}_{\Theta }+\frac{i\hbar c}{2}\overline{%
\Theta }\gamma ^{0}\left[ M^{-1},\partial _{0}M\right] \Theta ,  \label{a} \\
\mathcal{H}_{\Theta }& =\mathcal{H}_{\Psi }+\frac{i\hbar c}{2}\overline{\Psi 
}\gamma ^{0}\left[ M^{-1},\partial _{0}M^{-1}\right] \Psi .  \label{b}
\end{align}
It is noticeable that the equality $\mathcal{H}_{\Psi }=\mathcal{H}_{\Theta
} $ is valid at first order with respect to the $h_{\mu \nu }$'s. In fact,
since $\partial _{0}M$ and $\partial _{0}M^{-1}$ are first--order quantities,
this approximation is obtained by taking, for the first term in the
commutators in (\ref{a}) and (\ref{b}), the zeroth--order approximation of $%
M^{-1}$, that is the matrix unity.

\section{Interaction picture and the $S$--matrix}\index{Smatrix@$S$--matrix}

With the Lagrangian (\ref{lagtheta}) the interaction picture is easily
defined, since the corresponding conjugate momentum $\Pi _{\Theta }$ has the
same form as that of the free theory. This allows us to define the evolution
operator in that picture, then the $S$--matrix.

\subsection{Evolution operator and transition amplitudes}

The field in the interaction picture will be denoted by $\theta $. Let us
recall that this operator is obtained from the Heisenberg operator $\Theta $
by a unitary transformation such that the field equation becomes the free
one \cite{JR}. Accordingly, the evolution equation of the state vector reads 
\begin{equation}
i\hbar \frac{d}{dt}\left| \Phi (t)\right\rangle =H_{I}(t)\left| \Phi
(t)\right\rangle ,  \label{eqIP}
\end{equation}
where, in the absence of derivative coupling, which is the case for the
Lagrangian (\ref{lagtheta}), the Hamiltonian $H_{I}(t)$ is equal to the
interaction Hamiltonian expressed in terms of $\theta $. From the expression
(\ref{Htot}) of the total Hamiltonian one gets 
\begin{equation}
H_{I}(t)=\int (d^{3}x)\mathcal{H}_{\mathrm{int}}(x),  \label{HI}
\end{equation}
with 
\begin{equation}
\mathcal{H}_{\mathrm{int}} = -\frac{i\hbar c}{2}\overline{\theta }(\Gamma
^{k}-\gamma ^{k})(\partial _{k}\theta )+\frac{i\hbar c}{2}(\partial _{k}%
\overline{\theta })(\Gamma ^{k}-\gamma ^{k})\theta -\frac{\hbar c}{2}%
\overline{\theta }(\Gamma +2\frac{mc}{\hbar })\theta .  \label{Hint}
\end{equation}

The evolution operator in the interaction picture is, from (\ref{eqIP}), the
solution of the following equation together with the initial condition \cite
{SSS} : 
\begin{equation}
i\hbar \frac{d}{dt}U(t,t_{0})=H_{I}(t)U(t,t_{0})\,,\qquad U(t_{0},t_{0})=I\,.
\label{eqU}
\end{equation}
The perturbation theory is then obtained from the integral equation,
equivalent to (\ref{eqU}), 
\begin{equation}
U(t,t_{0})=I-\frac{i}{\hbar }\int_{t_{0}}^{t}H_{I}(\tau )U(\tau ,t_{0})d\tau
\,.
\end{equation}
In what follows, we are interested in the transition amplitudes to first
order with respect to the $h_{\mu \nu }$'s. These amplitudes will be
obtained from the first--order approximation with respect to the Hamiltonian $%
H_{I}$ of the \emph{U}--operator, namely 
\begin{equation}
U^{(1)}(t,t_{0})=-\frac{i}{\hbar }\int_{t_{0}}^{t}H_{I}(\tau )d\tau .
\end{equation}
By introducing the Hamiltonian density (\ref{Hint}) in normal form\footnote{%
See \cite{SSS}. To first order this prescription suppresses an infinite
contribution, due to the energy of the vacuum, in the transition amplitudes
for the antiparticles only. It can be seen, therefore, as an expression of
the symmetry between particles and antiparticles.}, and the initial and
final states of the transition, we will consider the amplitudes 
\begin{equation}
\left\langle \Phi _{f}\right| U^{(1)}(t,t_{0})\left| \Phi _{i}\right\rangle
=-\frac{i}{\hbar c}\int_{t_{0}}^{t}(d^{4}x)\left\langle \Phi _{f}|:\mathcal{H%
}_{\mathrm{int}}(x):|\Phi _{i}\right\rangle ,  \label{ampli}
\end{equation}
where the pair of colons $:\;\;:$ denotes as usual the normal product.

The first--order expressions needed for the evaluation of (\ref{ampli}) are
now derived from (\ref{M}) which gives 
\begin{equation}
\Lambda =M=I-\frac{h}{4}+\frac{1}{4}h^{0\mu }\gamma ^{0}\gamma _{\mu },
\end{equation}
then, from (\ref{gamk}) and (\ref{B_gam}), 
\begin{align}
\Gamma ^{k}-\gamma ^{k}& =\frac{1}{2}\left[ h^{00}\gamma ^{k}-h^{0k}\gamma
^{0}-h^{\mu k}\gamma _{\mu }\right]  \label{gamk1} \\
\Gamma +2\frac{mc}{\hbar }& =-\frac{mc}{\hbar }h^{00}+\frac{i}{4}(\partial
_{k}h^{0\mu })(\gamma ^{k}\gamma ^{0}\gamma _{\mu }-\gamma _{\mu }\gamma
^{0}\gamma ^{k}).  \label{gam1}
\end{align}
The corresponding expression for the Hamiltonian density $\mathcal{H}_{%
\mathrm{int}}$ is 
\begin{align}
\mathcal{H}_{\mathrm{int}}& =-\frac{i\hbar c}{4}\overline{\theta }\left[
h^{00}\gamma ^{k}-h^{0k}\gamma ^{0}-h^{\mu k}\gamma _{\mu }\right] (\partial
_{k}\theta )  \notag \\
& \quad +\frac{i\hbar c}{4}(\partial _{k}\overline{\theta })\left[
h^{00}\gamma ^{k}-h^{0k}\gamma ^{0}-h^{\mu k}\gamma _{\mu }\right] \theta 
\notag \\
& \quad +\frac{\hbar c}{2}\overline{\theta }\left[ \frac{mc}{\hbar }h^{00}-%
\frac{i}{4}(\partial _{k}h^{0\mu })(\gamma ^{k}\gamma ^{0}\gamma _{\mu
}-\gamma _{\mu }\gamma ^{0}\gamma ^{k})\right] \theta \,.  \label{hint}
\end{align}
In Section 5, the expression of the Hamiltonian $H_{I}(t)$ which will be
used is the expression obtained from the space integral of (\ref{hint}) by
performing the integration by parts of the term containing $\partial _{k}%
\overline{\theta }$, which gives 
\begin{equation}
H_{I}(t)=\int (d^{3}x)\,\overline{\theta (x)}\gamma ^{0}\mathcal{V}%
_{G}(x)\theta (x),  \label{hrho}
\end{equation}
where the operator $\mathcal{V}_{G}(x)$, acting on $\theta (x)$, is given by 
\begin{align}
\mathcal{V}_{G}(x)& =\frac{\hbar c}{2}\gamma ^{0}\left[ \frac{mc}{\hbar }%
h^{00}+\frac{i}{4}\partial _{k}h_{0j}\gamma ^{0}(\gamma ^{k}\gamma
^{j}-\gamma ^{j}\gamma ^{k})\right.  \notag \\
& \quad \left. +\frac{i}{2}(2\partial _{k}h^{0k}\gamma ^{0}+\partial
_{k}h^{jk}\gamma _{j}-\partial _{k}h^{00}\gamma ^{k})\right]  \notag \\
& +\frac{i\hbar c}{2}\gamma ^{0}\left[ 2h^{0k}\gamma ^{0}+h^{jk}\gamma
_{j}-h^{00}\gamma ^{k}\right] \partial _{k}\,.  \label{VG}
\end{align}
This form of the Hamiltonian is closely related to the equation of motion (%
\ref{eqtheta}) of the Heisenberg field $\Theta $. In fact, with the help of (%
\ref{gamk1}) and (\ref{gam1}), one checks that this equation can be written: 
\begin{equation}
\left( i\hbar c\gamma ^{\mu }\partial _{\mu }-mc^{2}\right) \Theta =\gamma
^{0}\,\mathcal{V}_{G}\Theta .  \label{dtki}
\end{equation}

In what follows, the initial and final states, which appear in (\ref{ampli}%
), are some one--particle or antiparticle states. They are defined from
positive or negative energy solutions $\chi _{i}$ and $\chi _{f}$ of the
free Dirac equation \index{Dirac equation} 
\begin{equation}
(i\gamma ^{\mu }\overrightarrow{\partial _{\mu }}-mc/\hbar )\chi
_{k}=0\,,\qquad \overline{\chi _{k}}(i\gamma ^{\mu }\overleftarrow{\partial
_{\mu }}+mc/\hbar )=0\,,\qquad k=i,f.
\end{equation}
Denoting by $\left| \chi _{k}\right\rangle $ the corresponding states, one
has then to calculate the matrix element 
\begin{equation}
\left\langle \chi _{f}\right| :\overline{\theta (x)}\gamma ^{0}\,\mathcal{V}%
_{G}(x)\theta (x):\left| \chi _{i}\right\rangle .  \label{matrel}
\end{equation}
As a free field, the operator $\theta $ can be written 
\begin{equation}
\theta (x)=\sum_{r=1}^{2}\int (d^{3}p)\left[ b_{r}(\vec{p})\chi _{\vec{p}%
,r}^{(+)}(x)+d_{r}^{\dag }(\vec{p})\chi _{\vec{p},r}^{(-)}(x)\right] \,,
\end{equation}
where $b_{r}(\vec{p})$ and $d_{r}(\vec{p})$ are the annihilation operators
for the particles or antiparticles, respectively, and $\chi _{\vec{p}%
,r}^{(\pm )}$ the positive or negative energy solutions of the free Dirac
equation \index{Dirac equation} given by \cite{BD,SSS} 
\begin{align}
\chi _{\vec{p},r}^{(+)}(x)& =\frac{1}{(2\pi \hbar )^{3/2}}\sqrt{\frac{mc^{2}%
}{E(\vec{p})}}u^{(r)}(\vec{p})e^{i(\vec{p}\cdot \vec{x}-E(\vec{p})t)/\hbar },
\label{kiplus} \\
\chi _{\vec{p},r}^{(-)}(x)& =\frac{1}{(2\pi \hbar )^{3/2}}\sqrt{\frac{mc^{2}%
}{E(\vec{p})}}v^{(r)}(\vec{p})e^{-i(\vec{p}\cdot \vec{x}-E(\vec{p})t)/\hbar
},  \label{kimoins}
\end{align}
with $E(\vec{p})=c\sqrt{p^{2}+m^{2}c^{2}}$, $p=\Vert \vec{p}\Vert $. In
terms of these, any solution $\chi $ with positive or negative energy can be
written 
\begin{equation}
\chi (x)=\sum_{r}\int (d^{3}p)\chi _{\vec{p},r}^{(\pm )}(x)(\chi _{\vec{p}%
,r}^{(\pm )},\chi )\,,  \label{B_chi}
\end{equation}
the scalar product of two solutions being defined by 
\begin{equation}
(\chi _{1},\chi _{2})=\int (d^{3}x)\overline{\chi _{1}}(x)\gamma ^{0}\chi
_{2}(x)\,.
\end{equation}
From (\ref{B_chi}) we have the following expression of the one--particle states 
\begin{equation}
\left| \chi \right\rangle =\sum_{r}\int (d^{3}p)(\chi _{\vec{p}%
,r}^{(+)},\chi )b_{r}^{\dag }(\vec{p})\left| \phi _{0}\right\rangle ,
\end{equation}
where $\left| \phi _{0}\right\rangle $ is the vacuum state, the
correspondence $\chi \rightarrow \left| \chi \right\rangle $ preserving the
scalar product. Denoting by $\theta ^{(+)}$ the positive frequency part of $%
\theta $, this last formula implies 
\begin{align}
\theta ^{(+)}(x)\left| \chi \right\rangle & =\sum_{r}\int (d^{3}p)\chi _{%
\vec{p},r}^{(+)}(x)b_{r}(\vec{p})\left| \chi \right\rangle  \notag \\
& =\sum_{r}\int (d^{3}p)\chi _{\vec{p},r}^{(+)}(x)(\chi _{\vec{p}%
,r}^{(+)},\chi )\left| \phi _{0}\right\rangle  \notag \\
& =\chi (x)\left| \phi _{0}\right\rangle .
\end{align}
The matrix element (\ref{matrel}) reduces to 
\begin{equation}
\left\langle \chi _{f}\right| \overline{\theta ^{(+)}(x)}\gamma ^{0}\,%
\mathcal{V}_{G}(x)\theta ^{(+)}(x)\left| \chi _{i}\right\rangle =\overline{%
\chi _{f}(x)}\gamma ^{0}\,\mathcal{V}_{G}(x)\chi _{i}(x),
\end{equation}
so that, from (\ref{hrho}), the amplitude (\ref{ampli}) reads 
\begin{equation}
\left\langle \chi _{f}\right| U^{(1)}(t,t_{0})\left| \chi _{i}\right\rangle
=-\frac{i}{\hbar c}\int_{t_{0}}^{t}(d^{4}x)\overline{\chi _{f}(x)}\gamma
^{0}\,\mathcal{V}_{G}(x)\chi _{i}(x).  \label{ampli2}
\end{equation}
This formula will be analyzed in more detail in Section 5.

\subsection{$S$--matrix}

The $S$--matrix can be defined when the external field vanishes at the limit
of infinite time\footnote{Here, this condition is realized, for example, in the case of a gravitational wave.}. It is obtained from the $U$ operator by taking the limits $t_{0}\rightarrow -\infty $ and $t\rightarrow +\infty $, and the
corresponding amplitudes can then be given from (\ref{ampli2}). However, it
is convenient here to return to the formula (\ref{ampli}) in which the
Hamiltonian density is given by (\ref{hint}). In fact, we want to show that
this expression can be transformed into a covariant one, the expression
already given by Gupta \cite{G2}.

The initial and final states being the same as in the calculation leading to
(\ref{ampli2}), formula (\ref{ampli}) with the expression (\ref{hint}) of $%
\mathcal{H}_{\mathrm{int}}$ gives, for the first--order $S$--matrix, 
\begin{align}
\left\langle \chi _{f}\right| S^{(1)}\left| \chi _{i}\right\rangle & =-i\int
(d^{4}x)\left\{ \frac{1}{2}\overline{\chi _{f}}\left[ \frac{mc}{\hbar }%
h^{00}-\frac{i}{4}(\partial _{k}h^{0\mu })(\gamma ^{k}\gamma ^{0}\gamma
_{\mu }-\gamma _{\mu }\gamma ^{0}\gamma ^{k})\right] \chi _{i}\right.  \notag
\\
& \quad +\frac{i}{4}\overline{\chi _{f}}\left[ h^{\mu k}\gamma _{\mu
}+h^{0k}\gamma ^{0}-h^{00}\gamma ^{k}\right] (\partial _{k}\chi _{i})  \notag
\\
& \quad \left. -\frac{i}{4}(\partial _{k}\overline{\chi _{f}})\left[ h^{\mu
k}\gamma _{\mu }+h^{0k}\gamma ^{0}-h^{00}\gamma ^{k}\right] \chi
_{i}\right\} \,.  \label{S1}
\end{align}
This expression can be simplified by integrating by parts the term
containing the derivative $\partial _{k}h^{0\mu }$. This leads to the two
terms 
\begin{equation*}
\frac{i}{8}h^{0\mu }\left[ (\partial _{k}\overline{\chi _{f}})(\gamma
^{k}\gamma ^{0}\gamma _{\mu }-\gamma _{\mu }\gamma ^{0}\gamma ^{k})\chi _{i}+%
\overline{\chi _{f}}(\gamma ^{k}\gamma ^{0}\gamma _{\mu }-\gamma _{\mu
}\gamma ^{0}\gamma ^{k})(\partial _{k}\chi _{i})\right] .
\end{equation*}
In each of these, by introducing either of the formulas 
\begin{eqnarray}
\gamma ^{k}\gamma ^{0}\gamma _{\mu }-\gamma _{\mu }\gamma ^{0}\gamma ^{k}
&=&2[\delta _{\mu }^{0}\gamma ^{k}-\delta _{\mu }^{k}\gamma ^{0}-\gamma
_{\mu }\gamma ^{0}\gamma ^{k}]  \notag \\
&=&2[\delta _{\mu }^{k}\gamma ^{0}-\delta _{\mu }^{0}\gamma ^{k}+\gamma
^{k}\gamma ^{0}\gamma _{\mu }]\,,
\end{eqnarray}
one can insert the derivatives in the combinations $\gamma ^{k}\partial
_{k}\chi _{i}$ or $\partial _{k}\overline{\chi _{f}}\gamma ^{k}$, yielding 
\begin{equation}
\frac{i}{4}h^{0k}\overline{\chi _{f}}\gamma ^{0}(\overleftarrow{\partial _{k}%
}-\overrightarrow{\partial _{k}})\chi _{i}+\frac{i}{4}h^{0\mu }\left[ 
\overline{\chi _{f}}(\delta _{\mu }^{0}-\gamma _{\mu }\gamma ^{0})(\gamma
^{k}\partial _{k}\chi _{i})-(\partial _{k}\overline{\chi _{f}}\gamma
^{k})(\delta _{\mu }^{0}-\gamma ^{0}\gamma _{\mu })\chi _{i}\right] \,.
\end{equation}
Adding this contribution to the remaining terms in (\ref{S1}), one gets 
\begin{eqnarray}
\left\langle \chi _{f}\right| S^{(1)}\left| \chi _{i}\right\rangle &=&-i\int
(d^{4}x)\left\{ \frac{mc}{2\hbar }h^{00}\overline{\chi _{f}}\chi _{i}+\frac{i%
}{4}h^{\mu k}\overline{\chi _{f}}\gamma _{\mu }(\overrightarrow{\partial _{k}%
}-\overleftarrow{\partial _{k}})\chi _{i}\right.  \notag \\
&&\qquad \left. -\frac{i}{4}h^{\mu 0}\left[ \overline{\chi _{f}}\gamma _{\mu
}\gamma ^{0}(\gamma ^{k}\partial _{k}\chi _{i})-(\partial _{k}\overline{\chi
_{f}}\gamma ^{k})\gamma ^{0}\gamma _{\mu }\chi _{i}\right] \right\} \,.
\end{eqnarray}
Finally, using the Dirac equation in the last bracket, this expression
reduces to 
\begin{equation}
\left\langle \chi _{f}\right| S^{(1)}\left| \chi _{i}\right\rangle =-i\int
(d^{4}x)\frac{i}{4}h^{\mu \nu }\overline{\chi _{f}}\left( \gamma _{\mu }%
\overrightarrow{\partial _{\nu }}-\overleftarrow{\partial _{\nu }}\gamma
_{\mu }\right) \chi _{i}.  \label{S2}
\end{equation}
This formula agrees with the rule given by Gupta \cite{G2} : the expression
appearing under the integral sign is, up to the sign, obtained from the
interaction Lagrangian $\mathcal{L}-\mathcal{L}_{0}$ in (\ref{Lag}) by
replacing $\Psi $ by $\chi _{i}$ and $\overline{\Psi }$ by $\overline{\chi
_{f}}$\thinspace \footnote{%
The contribution of the term $(h/2)\mathcal{L}_{0}$ of (\ref{Lag}) vanishes
in the first order considered here, since $\chi _{i}$ and $\chi _{f}$ are
solutions of the free Dirac equation. But this property is limited to the
first order.}. This result is identical with the one valid in the case of a
nonderivative coupling, where the interaction picture exists directly. More
generally, the validity of the rule asserted by Gupta, implying the use of
the interaction Lagrangian defined by (\ref{Lag}) and the covariant form of
the propagators, can be proved at higher orders in the quantized theory
defined in Section 3.

\section{Calculation of the relativistic phase shifts in the weak--field
approximation}

In this final section we use the tools and the material derived in the
previous sections to make an explicit calculation of the various
contributions to a gravitationally induced phase shift in matter--wave
interferometry. We restrict ourselves to one--particle or one--antiparticle
states. The application of the formalism to many--particle states and
coherent beams of massive particles of different spins will be developed in
another publication. We assume that the incoming particles or antiparticles
are described by the state vector $\left| \chi (t_{0})\right\rangle =\left|
\chi _{i}\right\rangle $ at some time $t_{0}$ before interaction ($t_{0}$
can be conveniently taken to be $-\infty $). At a later time $t$, this state
evolves into $\left| \chi (t)\right\rangle $ which interferes with a
reference beam described by $\left| \chi _{\mathrm{ref}}\right\rangle $ with
which it is recombined in the final beam splitter. In practice, $\left| \chi
_{\mathrm{ref}}\right\rangle $ is produced by the other arm of the
interferometer and in many cases, one will have $\left| \chi _{\mathrm{ref}%
}\right\rangle \equiv \left| \chi _{i}\right\rangle .$

We are thus interested in the spinorial wave function for one--(anti)particle
states: 
\begin{equation}
\chi (x)=\left\langle \phi _{0}\right| \theta (x)\left| \chi (t)\right\rangle
\label{spinor}
\end{equation}
where \ $\theta (x)$, $\ \left| \chi (t)\right\rangle $, $\left| \phi
_{0}\right\rangle $ are respectively the free--field operator, the
one--(anti)particle and the vacuum state vectors in the interaction
representation. It is easily shown that, when pair creations are neglected,
this expression is equivalent to the Heisenberg field amplitude $%
\left\langle 0,in\right| \Theta (x) \left| \Phi \right\rangle $.

The interference signal itself is given by the projection: 
\begin{eqnarray}
\int d^{3}x\,\chi _{\mathrm{ref}}^{\dagger }(x)\chi (x) &=&\int
d^{3}x\left\langle \chi _{\mathrm{ref}}\right| \theta ^{\dag }(x)\left| \phi
_{0}\right\rangle \left\langle \phi _{0}\right| \theta (x)\left| \chi
(t)\right\rangle  \notag \\
&=&\left\langle \chi _{\mathrm{ref}}\right| U(t,t_{0})\left| \chi
(t_{0})\right\rangle
\end{eqnarray}
where the space integral is over some detection volume. More generally one
should consider a detection hypersurface $\sigma (x)$ and the projection: 
\begin{equation}
\int_{\sigma }d\sigma ^{\mu }\,\overline{\chi }_{\mathrm{ref}}(x)\gamma
_{\mu }\chi (x)=\int_{\sigma }d\sigma ^{\mu }\;\left\langle \chi _{%
\mathrm{ref}}\right| \overline{\theta }(x)\left| \phi _{0}\right\rangle
\gamma _{\mu }\left\langle \phi _{0}\right| \theta (x)\left| \chi
(t)\right\rangle \,.
\end{equation}
In this paper, we shall limit ourselves to the calculation of the amplitude: 
\begin{equation}
\left\langle \chi _{\mathrm{ref}}\right| \chi (t)\rangle =\left\langle \chi
_{\mathrm{ref}}\right| U(t,t_{0})\left| \chi (t_{0})\right\rangle \,.
\end{equation}
and take the phase of this complex amplitude as the phase contribution of
the perturbing interaction. The resulting spinor (\ref{spinor}) will also be
derived using two different methods: first, in configuration space and
second, in the momentum representation.

\subsection{Calculation in configuration space}

The evolution equation of the state vector in the interaction picture is 
\begin{equation}
i\hbar \frac{d}{dt}\left| \chi (t)\right\rangle =H_{I}(t)\left| \chi
(t)\right\rangle ,  \label{TS}
\end{equation}
where the Hamiltonian $H_{I}(t)$ is 
\begin{equation}
H_{I}(t)=\int d^{3}x\,\;\theta ^{\dag }(x)\mathcal{V}_{G}(x)\theta (x)\;,
\end{equation}
and where the operator $\mathcal{V}_{G}(x)$, acting on the field operator $%
\theta (x)$, is given to first order by (\ref{VG}) 
\begin{eqnarray}
\mathcal{V}_{G} &=&\frac{1}{2}mc^{2}\gamma ^{0}h^{00}+\frac{i\hbar c}{8}%
\partial _{k}h_{0j}(\gamma ^{k}\gamma ^{j}-\gamma ^{j}\gamma ^{k})  \notag \\
&&+\frac{i\hbar c}{4}\gamma ^{0}(2\partial _{k}h^{0k}\gamma ^{0}+\partial
_{k}h^{jk}\gamma _{j}-\partial _{k}h^{00}\gamma ^{k})  \notag \\
&&+\frac{i\hbar c}{2}\gamma ^{0}\left[ 2h^{0k}\gamma ^{0}+h^{jk}\gamma
_{j}-h^{00}\gamma ^{k}\right] \partial _{k}
\end{eqnarray}
that we shall write: 
\begin{equation}
\mathcal{V}_{G}(x)=A(x)+\frac{i\hbar }{2}\partial _{j}B^{j}(x)+i\hbar
B^{j}(x)\partial _{j}=A(x)+\frac{1}{2}\left\{ i\hbar \partial
_{j},B^{j}(x)\right\} _{+}
\end{equation}
with: 
\begin{eqnarray}
A(x) &=&\frac{1}{2}mc^{2}\gamma ^{0}h^{00}+\frac{\hbar c}{4}\sigma
^{kj}\partial _{k}h_{0j}  \notag \\
B^{j}(x) &=&\frac{c}{2}\gamma ^{0}\left[ 2h^{0j}\gamma ^{0}+h^{kj}\gamma
_{k}-h^{00}\gamma ^{j}\right] \ .
\end{eqnarray}
From equation (\ref{spinor}) we check that the evolution of the
one-(anti)particle spinor is governed by the equation: 
\begin{equation}
i\hbar \partial _{t}\chi =-i\hbar c\gamma ^{0}\gamma ^{j}\partial _{j}\chi
+mc^{2}\gamma ^{0}\chi +\mathcal{V}_{G}\chi  \label{dtkhi}
\end{equation}
to which we may add terms corresponding to diagonal magnetic dipole and
off--diagonal electric dipole interactions \cite{Borde6,BKT}. This equation
has been used in references \cite{Borde6} and \cite{BKT} to discuss all the
terms that lead to a phase shift in an interferometer.

To obtain the corresponding amplitude, we can start directly from the
integral form of (\ref{TS}): 
\begin{equation}
\left| \chi (t)\right\rangle =\left| \chi (t_{0})\right\rangle -\frac{i}{%
\hbar c}\int_{t_{0}}^{t}d^{4}x^{\prime }\theta ^{\dagger }(x^{\prime })%
\mathcal{V}_{G}(x^{\prime })\theta (x^{\prime })\left| \chi (t^{\prime
})\right\rangle  \label{integral}
\end{equation}
So that, to first order: 
\begin{eqnarray}
\!\!\!\!\!\!\!\! \left\langle \chi _{\mathrm{ref}}\right| \chi ^{(1)}(t)\rangle
&=&\left\langle \chi _{\mathrm{ref}}\right| U^{(1)}(t,t_{0})\left| \chi
_{i}\right\rangle  \notag \\
&=&-\frac{i}{\hbar c}\int_{t_{0}}^t d^4 x^\prime \chi_{\mathrm{ref}}^{\dagger }(x^{\prime })\mathcal{V}_{G}(x^{\prime })\chi _{i}(x^{\prime }) 
\notag \\
&=&-\frac{i}{\hbar c}\int_{t_{0}}^t d^4 x \left\{\chi_{\mathrm{ref}}^{\dagger }\left[ A(x)+\frac{i\hbar}{2}\partial _{j}B^{j}(x)+i\hbar
B^{j}(x)\partial_j\right] \chi _{i}\right\} \notag \\
&=&\int_{t_0}^t d^4 x \left\{\chi_{\mathrm{ref}}^{\dagger}\left[ 
\frac{h^{00}}{2c}\gamma ^{0}\partial_t + \vec{h} \cdot \vec{\nabla}-\frac{1}{2} \vec{\alpha} \cdot \overset{\Rightarrow }{h} \cdot \vec{\nabla}\right. \right.  \notag \\
&&\left. \left. +\frac{i}{4}\vec{\Sigma} \cdot \vec{\nabla}\times \vec{h}-\frac{1}{4}\left(\vec{\nabla} h^{00}+\vec{\nabla} \cdot \overset{\Rightarrow }{h}\right) \cdot \vec{\alpha}+\frac{1}{2}\vec{\nabla} \cdot \vec{h}\right] \chi _{i}\right\}
\label{PSCS}
\end{eqnarray}
which follows also from Gupta's form. We have used the definitions: 
\begin{equation}
\vec{\alpha}=\gamma ^{0}\vec{\gamma}=\left( 
\begin{array}{cc}
0 & \vec{\sigma} \\ 
\vec{\sigma} & 0
\end{array}
\right) ,\vec{\Sigma}=\left( 
\begin{array}{cc}
\vec{\sigma} & 0 \\ 
0 & \vec{\sigma}
\end{array}
\right) ,\vec{h}=\{h^{0k}\},\overset{\Rightarrow }{h}=\{h^{ij}\} \, .
\end{equation}
The calculation of the spinor itself, by a stationary phase method in
configuration space, is outlined in Appendix C.

In equation (\ref{PSCS}), the first three terms lead to the familiar phase
shifts of the Linet--Tourrenc formula \cite{LT}, the fourth term is the
spin--rotation interaction and the last two terms ensure hermiticity. But,
because of the Dirac matrices, the interpretation of the various terms in
configuration space is not so transparent and, in previous works
non--relativistic limits have been taken either directly as in our own work 
\cite{BKT} or through a Foldy--Wouthuysen transformation as in \cite{19} in
the case of inertial fields, to put equations (\ref{dtkhi}) and (\ref{PSCS})
in a form where the significance of the terms is more obvious.

In this paper, we may as well take advantage of the flat Minkowski
space--time and, therefore, we will use rather the momentum representation in
the following. As we shall see, the interpretation of the terms is then much
easier, even in their relativistic form.

\subsection{Calculation in the momentum representation}

The free--field operator $\theta $ is written as before as: 
\begin{equation}
\theta (x)=\sum_{r=1}^{2}\int d^{3}p \left[ b_{r}(\vec{p})\chi_{\vec{p} ,r}^{(+)}(x)+d_{r}^{\dag }(\vec{p})\chi _{\vec{p},r}^{(-)}(x)\right] ,
\end{equation}
where $b_{r}(\vec{p})$ and $d_{r}(\vec{p})$ are the annihilation operators
for the particles or antiparticles, respectively, and $\chi _{\vec{p} ,r}^{(\pm )}$ are the positive or negative energy solutions of the free
Dirac equation given by (\ref{kiplus}) and (\ref{kimoins}).

Let us introduce the Fourier transforms $\widetilde{A}(\vec{k},t)$, $\widetilde{B}^{j}(\vec{k},t)$, $\widetilde{h}_{\mu \nu }(\vec{k})$: 
\begin{align}
A(x)& =\frac{1}{(2\pi )^{3/2}}\int d^{3}k \;\widetilde{A}(\vec{k},t)e^{i\vec{k%
}\cdot \vec{x}} \\
B^{j}(x)& =\frac{1}{(2\pi )^{3/2}}\int d^3 k\; \widetilde{B}^{j}(\vec{k}%
,t)e^{i\vec{k}\cdot \vec{x}} \\
h_{\mu \nu }(x)& =\frac{1}{(2\pi )^{3/2}}\int d^{3}k \; \widetilde{h}_{\mu \nu
}(\vec{k},t)e^{i\vec{k}\cdot \vec{x}} \\
\mathcal{V}_{G}(x)e^{i\vec{p}\cdot \vec{x}/\hbar }& =\frac{1}{(2\pi )^{3/2}} \int d^{3}k \;\widetilde{\mathcal{V}}_{G}(\vec{k},\vec{p},t)e^{i\vec{k}\cdot 
\vec{x}}e^{i\vec{p}\cdot \vec{x}/\hbar } \\
\widetilde{\mathcal{V}_{G}}(\vec{k},\vec{p},t)& =\widetilde{A}(\vec{k},t)- \widetilde{\vec{B}}(\vec{k},t)\cdot \left( \vec{p}+{\textstyle\frac{1}{2}} \hbar \vec{k}\right) \, .
\end{align}
Let us expand $\chi (x)$, $\overline{\chi _{\mathrm{ref}}}(x)$ and $\chi
_{i}(x)$ in plane waves using the expansions of $\ \overline{\theta ^{(\pm )}%
}(x)$ and $\theta ^{(\pm )}(x)$. For particles, the output spinor is: 
\begin{eqnarray}
\chi (x) &=&\left\langle \phi _{0}\right| \theta ^{(+)}(x)\left| \chi
(t)\right\rangle  \notag \\
&=&\left\langle \phi _{0}\right| \sum_{r=1}^{2}\int d^3 p \; b_{r}(\vec{p}%
)\chi _{\vec{p},r}^{(+)}(x)\left| \chi (t)\right\rangle  \notag \\
&=&\frac{1}{(2\pi \hbar )^{3/2}}\sum_{r=1}^{2}\int d^3 p\sqrt{\frac{mc^{2}%
}{E(\vec{p})}}u^{(r)}(\vec{p})e^{i(\vec{p}\cdot \vec{x}-E(\vec{p})t)/\hbar
}\langle 1_{\vec{p},r}\left| \chi (t)\right\rangle \qquad
\end{eqnarray}
with $\left| 1_{\vec{p},r}\right\rangle =$ $b_{r}^{\dagger }(\vec{p})|\phi
_{0}\rangle $ and with, using again (\ref{integral}), 
\begin{eqnarray}
\left\langle \chi _{\mathrm{ref}}\right| \chi (t)\rangle & = & \left\langle
\chi _{\mathrm{ref}}\right| U(t,t_{0})\left| \chi _{i}\right\rangle  \notag
\\
& = & \langle \chi _{\mathrm{ref}}\left| \chi _{i}\right\rangle -\frac{i}{\hbar
c}\int_{t_{0}}^{t} d^4 x^\prime \chi _{\mathrm{ref}}^{\dagger
}(x^{\prime })\mathcal{V}_{G}(x^{\prime })\chi _{i}(x^{\prime }) \\
& = & \langle \chi _{\mathrm{ref}}\left| \chi _{i}\right\rangle -\frac{i}{\hbar 
}\frac{1}{(2\pi \hbar )^{3}}\sum_{r,r^{\prime }}\int_{t_0}^t dt^{\prime
}\int d^3 x^\prime \; \int d^3 p \; d^3 p^{\prime } \langle \chi _{\mathrm{ref}}\left| 1_{\vec{p},r}\right\rangle  \notag \\
& & \sqrt{\frac{%
mc^{2}}{E(\vec{p})}}u^{(r)\dagger }(\vec{p})e^{-i(\vec{p}.\vec{x}^{\prime
}-E(\vec{p}) t^\prime)/\hbar} \frac{1}{(2\pi)^{3/2}}\int d^3 k \; \widetilde{\mathcal{V}}_{G}(\vec{k},%
\vec{p}^{\prime },t^{\prime })e^{i\vec{k}\cdot \vec{x}^{\prime }}  \notag \\
&&\sqrt{\frac{mc^{2}}{E(\vec{p}^{\prime })}}u^{(r^{\prime })}(\vec{p}%
^{\prime })e^{i(\vec{p}^{\prime }\cdot \vec{x}^{\prime }-E(\vec{p}^{\prime
})t^{\prime })/\hbar }\langle 1_{\vec{p}^{\prime },r^{\prime }}\left| \chi
_{i}\right\rangle \\
&=&\langle \chi _{\mathrm{ref}}\left| \chi _{i}\right\rangle -\frac{i}{\hbar 
}\sum_{r,r^{\prime }}\int_{t_0}^t dt^\prime \int d^3 p\int \frac{%
d^{3}k}{(2\pi )^{3/2}}\langle \chi _{\mathrm{ref}}\left| 1_{\vec{p}+\hbar 
\vec{k},r}\right\rangle  \notag \\
&&\sqrt{\frac{mc^{2}}{E(\vec{p}+\hbar \vec{k})}}\sqrt{\frac{mc^{2}}{E(\vec{p}%
)}}u^{(r)\dagger }(\vec{p}+\hbar \vec{k})\widetilde{\mathcal{V}}_{G}(\vec{k},%
\vec{p},t^{\prime })u^{(r^{\prime })}(\vec{p})  \notag \\
&&e^{i\left[ E(\vec{p}+\hbar \vec{k})-E(\vec{p})\right] t^{\prime }/\hbar
}\langle 1_{\vec{p},r^{\prime }}\left| \chi (t^{\prime })\right\rangle
\label{scat}
\end{eqnarray}

\landscape
\noindent and similar expressions for antiparticles. In many cases $\left| \chi
_{i}\right\rangle $ and $\left| \chi _{\mathrm{ref}}\right\rangle $ can be
conveniently taken as plane waves, but it is usually more interesting to
consider wave packets. Replacing $\left| \chi _{\mathrm{ref}}\right\rangle $
by $\left| 1_{\vec{p},r}\right\rangle $, we check that the momentum
representation $\langle 1_{\vec{p},r}\left| \chi (t)\right\rangle $ of $\chi
(x)$ satisfies: 
\begin{eqnarray}
\lefteqn{i\hbar \partial _{t}\langle \phi _{0}|b_{r}(\vec{p})|\chi (t)\rangle} \nonumber \\ &=&\sum_{r^{\prime }}\int \frac{d^{3}k}{\left( 2\pi \right) ^{3/2}} \sqrt{\frac{mc^{2}}{E(\vec{p})}}\sqrt{\frac{mc^{2}}{E(\vec{p}-\hbar \vec{k})}} u^{(r)\dagger }(\vec{p})\widetilde{\mathcal{V}}_{G}(\vec{k},\vec{p}-\hbar 
\vec{k},t)u^{(r^{\prime })}(\vec{p}-\hbar \vec{k}) e^{i\left[ E(\vec{p})-E(\vec{p}-\hbar \vec{k})\right] t/\hbar }\langle
\phi _{0}|b_{r^{\prime }}(\vec{p}-\hbar \vec{k})|\chi (t)\rangle \, .
\label{eqmomentum}
\end{eqnarray}
This equation is, in momentum representation, the analogous of equation (\ref
{dtkhi}) in configuration space and we shall give its first--order solution
later. It leads to a discrete set of coupled equations for a fixed or
negligible recoil momentum. We illustrate below, in the case of the
scattering amplitude, how the matrix element, which appears in the second
member of (\ref{scat}) and (\ref{eqmomentum}) can be evaluated.

To first order, the scattering amplitude (\ref{scat}) is: 
\begin{eqnarray}
\langle \chi _{\mathrm{ref}}|\chi ^{(1)}(t)\rangle &=&\langle \chi _{\mathrm{%
ref}}|U^{(1)}(t,t_{0})|\chi _{i}\rangle \nonumber\\
& = & -\frac{i}{\hbar }\sum_{r,r^{\prime }}\int_{t_{0}}^{t}dt^{\prime }\int
d^{3}p \int \frac{d^{3}k}{(2\pi )^{3/2}}\langle \chi _{\mathrm{ref}}|1_{%
\vec{p}+\hbar \vec{k},r}\rangle \sqrt{\frac{mc^{2}}{E(\vec{p}+\hbar \vec{k})}}\sqrt{\frac{mc^{2}}{E(\vec{p})}} \nonumber\\
& & \qquad\qquad\qquad\qquad\qquad u^{(r)\dagger }(\vec{p}+\hbar \vec{k})\widetilde{\mathcal{V}}_{G}(\vec{k},%
\vec{p},t^{\prime })u^{(r^{\prime })}(\vec{p}) e^{i\left[ E(\vec{p}+\hbar \vec{k})-E(\vec{p})\right] t^{\prime }/\hbar
}\langle 1_{\vec{p},r^{\prime }}\left| \chi _{i}\right\rangle \, .
\label{amplitude}
\end{eqnarray}

The next step takes benefit from the smallness of $\hbar \vec{k}/mc$ or of $%
\hbar \vec{k}c/E(\vec{p})$ to expand the various quantities in this
expression to first order in these parameters. The energy $E(\vec{p}+\hbar 
\vec{k})$ can be expanded in a Taylor series: 
\begin{equation}
E(\vec{p}+\hbar \vec{k})=E(\vec{p})+\frac{\hbar \vec{k}\cdot \vec{p}c^{2}}{E(%
\vec{p})}+\frac{(\hbar k)^{2}c^{2}}{2E(\vec{p})}+\ldots =E(\vec{p})+\hbar 
\vec{k}\cdot \vec{v}+\hbar \delta +\ldots \, , \label{energy}
\end{equation}
where $\delta $ is the recoil shift. From the general transformation law of
spinors in Lorentz boosts (see for example \cite{LL2}): 
\begin{equation}
u(\vec{p})=\left[ \cosh \left( \frac{\varphi }{2}\right) +\widehat{\vec{n}}%
\cdot \vec{\alpha}\sinh \left( \frac{\varphi }{2}\right) \right] u(0) \, ,
\end{equation}
where $\widehat{\vec{n}}$ is the unit vector along $\vec{p}$, and $\tanh
\varphi =\left| \vec{p}\right| /(\gamma mc)$, we derive the following
infinitesimal transformation for spinors: 
\begin{eqnarray}
u(\vec{p}+\delta \vec{p}) &=&\left\{ 1+\frac{\widehat{\vec{n}}\cdot \vec{%
\alpha}}{2}\tanh \varphi \frac{\widehat{\vec{n}}\cdot \delta \vec{p}}{p} +\frac{1}{2}\sinh \varphi \left[ \frac{\vec{\alpha}\cdot \delta 
\vec{p}}{p}-\left( \widehat{\vec{n}}\cdot \vec{\alpha}\right) \frac{\widehat{%
\vec{n}}\cdot \delta \vec{p}}{p}\right] +i\sinh ^{2}\left( \frac{\varphi }{2}%
\right) \frac{\widehat{\vec{n}}\times \delta \vec{p}}{p}\cdot \vec{\Sigma}%
\right\} u(\vec{p}) 
\end{eqnarray}
we have to first order in $\hbar \vec{k}/mc$: 
\begin{equation}
u^{\dagger }(\vec{p}+\hbar \vec{k})=u^{\dagger }(\vec{p})\{1+\frac{1}{%
2\gamma }\frac{\hbar \vec{k}}{mc}\cdot \vec{\alpha}-\frac{i}{2(\gamma +1)}%
\frac{\vec{p}}{mc}\times \frac{\hbar \vec{k}}{mc}\cdot \vec{\Sigma}\}\,,
\end{equation}
where $\gamma =1/\sqrt{1-\beta ^{2}}$, $\vec{k}=\vec{k}_{\Vert }+\gamma \vec{%
k}_{\bot }$, the indices $\Vert $ and $\bot $ designate vector parts
respectively parallel and perpendicular to $\vec{p}$. The term proportional
to $\vec{\alpha}$ represents a boost (velocity change) and the term
proportional to $\vec{\Sigma}$ \ a rotation \index{Thomas precession} (Thomas precession).

Rather than to calculate directly 
\begin{equation}
\sqrt{\frac{mc^{2}}{E(\vec{p} + \hbar \vec{k})}} \sqrt{\frac{mc^{2}}{E(\vec{p})}} u^{(r)\dagger }(\vec{p}+\hbar \vec{k}) \widetilde{\mathcal{V}}_{G}(\vec{k},\vec{p},t)u^{(r^{\prime })}(\vec{p})
\end{equation}
it is simpler to calculate first the matrix element: 
\begin{eqnarray}
\lefteqn{\sqrt{\frac{mc^{2}}{E(\vec{p}+ {\textstyle \frac{1}{2}} \hbar \vec{k})}}\sqrt{\frac{mc^{2}}{E(\vec{p}- {\textstyle \frac{1}{2}} \hbar \vec{k})}}u^{(r)\dag }(\vec{p}+ {\textstyle \frac{1}{2}} \hbar \vec{k})\widetilde{\mathcal{V}_{G}}(\vec{k},\vec{p}- {\textstyle \frac{1}{2}} \hbar \vec{k},t)u^{(r^{\prime })}(\vec{p} - {\textstyle \frac{1}{2}} \hbar \vec{k})} \notag \\
\qquad\qquad &=&\frac{mc^{2}}{E(\vec{p})}u^{(r)\dag }(\vec{p}+ {\textstyle \frac{1}{2}} \hbar \vec{k})\left[ 
\widetilde{A}(\vec{k},t)-\widetilde{\vec{B}}(\vec{k},t)\cdot \vec{p}\right]
u^{(r^{\prime })}(\vec{p}- {\textstyle \frac{1}{2}} \hbar \vec{k}) \\
\qquad\qquad &=&\frac{mc^{2}}{E(\vec{p})}u^{(r)\dag }(\vec{p})\left[ \widetilde{A}(\vec{k}%
,t)-\widetilde{\vec{B}}(\vec{k},t)\cdot \vec{p}\right] u^{(r^{\prime })}(\vec{p}) +\frac{1}{4\gamma }\frac{mc^{2}}{E(\vec{p})}u^{(r)\dag }(\vec{p})\left\{ 
\frac{\hbar \vec{k}}{mc}\cdot \vec{\alpha},\widetilde{A}(\vec{k},t)- \widetilde{\vec{B}}(\vec{k},t)\cdot \vec{p}\right\} _{-}u^{(r^{\prime })}(\vec{p}) \notag\\ 
\qquad\qquad & & -\frac{i}{4(\gamma +1)}\frac{mc^{2}}{E(\vec{p})}u^{(r)\dag }(\vec{p})\left\{ \frac{\vec{p}}{mc}\times \frac{\hbar \vec{k}}{mc}\cdot \vec{\Sigma},\widetilde{A}(\vec{k},t)-\widetilde{\vec{B}}(\vec{k},t)\cdot \vec{p}\right\}
_{+}u^{(r^{\prime })}(\vec{p}) \, ,
\end{eqnarray}
where $\left\{ A,B\right\} _{\pm }$ designate (anti)commutators. The first
line gives: 
\begin{eqnarray}
\lefteqn{\frac{mc^{2}}{E(\vec{p})}u^{(r)\dag }(\vec{p})\left[ \widetilde{A}(\vec{k},t)-\widetilde{\vec{B}}(\vec{k},t)\cdot \vec{p}\right] u^{(r^{\prime })}(\vec{p})} \notag \\
\qquad\qquad &=&\left[ \frac{E(\vec{p})\widetilde{h}_{00}}{2}-c\vec{p}\cdot \widetilde{%
\vec{h}}+\frac{c^{2}}{2E(\vec{p})}\vec{p}\cdot \widetilde{\overset{%
\Rightarrow }{h}}\cdot \vec{p}\right] \delta _{rr^{\prime }} -\frac{i\hbar c}{4\gamma }(\vec{k}\times \widetilde{\vec{h}})\cdot
w^{(r)\dag }(\vec{\sigma}_{\bot }+\gamma \vec{\sigma}_{\Vert })w^{(r^{\prime
})}  \notag \\
\qquad\qquad &=&\frac{c^{2}}{2E(\vec{p})}p^{\mu }\widetilde{h}_{\mu \nu }p^{\nu }\delta
_{rr^{\prime }}-\frac{i\hbar c}{4\gamma }(\vec{k}\times \widetilde{\vec{h}}%
)\cdot w^{(r)\dag }\vec{a}w^{(r^{\prime })} \, ,
\end{eqnarray}
where $w^{(r)}$ are Pauli two-component spinors corresponding either to
helicity eigenvalues or to the two values of the $z$-component of the spin
in the rest frame, and where 
\begin{equation}
\vec{a}=(\vec{\sigma}_{\bot }+\gamma \vec{\sigma}_{\Vert })
\end{equation}
is the spatial part of the Thomas--Pauli--Lubanski 4--vector \index{Pauli--Lubanski vector} operator \cite{Thomas,PL}. The second line gives the term: 
\begin{equation}
\frac{1}{4\gamma }\frac{mc^{2}}{E(\vec{p})}u^{(r)\dag }(\vec{p})\left\{ 
\frac{\hbar \vec{k}}{mc}\cdot \vec{\alpha},\widetilde{A}(\vec{k},t)- \widetilde{\vec{B}}(\vec{k},t)\cdot \vec{p}\right\} _{-}u^{(r^{\prime })}(\vec{p}) =\frac{i\hbar c^{2}}{4E(\vec{p})\gamma }(\vec{k}\times \widetilde{\overset{\Rightarrow }{h}}\cdot \vec{p})\cdot w^{(r)\dag }\vec{a}w^{(r^{\prime })}
\end{equation}
The last line gives the Thomas precession terms: 
\begin{eqnarray}
\lefteqn{-\frac{i}{4(\gamma +1)}\frac{mc^{2}}{E(\vec{p})}u^{(r)\dag }(\vec{p}%
)\left\{ \frac{\vec{p}}{mc}\times \frac{\hbar \vec{k}}{mc}\cdot \vec{\Sigma}%
\;,\;\widetilde{A}(\vec{k},t)-\widetilde{\vec{B}}(\vec{k},t)\cdot \vec{p}%
\right\} _{+}u^{(r^{\prime })}(\vec{p})}  \\
\quad &=&\frac{i\hbar \widetilde{h}_{00}}{4m(\gamma +1)}(\vec{k}\times \vec{p}%
)\cdot w^{(r)\dag }\vec{a}w^{(r^{\prime })} -\frac{i\hbar c}{2m(\gamma +1)}\frac{\vec{p}\cdot \widetilde{\vec{h}}}{E(\vec{p})}(\vec{k}\times \vec{p})\cdot w^{(r)\dag }\vec{a}w^{(r^{\prime })} 
+\frac{i\hbar }{4E^{2}(\vec{p})m(\gamma +1)}\left[ (\vec{k}\times \vec{p}%
)\cdot \widetilde{\overset{\Rightarrow }{h}}\cdot \vec{p}\right] \vec{p}%
\cdot w^{(r)\dag }\vec{a}w^{(r^{\prime })}  \notag
\end{eqnarray}
The last line can be rewritten to yield: 
\begin{eqnarray}
\!\!\!\!\!\!\!\!\lefteqn{\sqrt{\frac{mc^{2}}{E(\vec{p}+ {\textstyle \frac{1}{2}} \hbar \vec{k})}}\sqrt{\frac{mc^{2}}{E(\vec{p}- {\textstyle \frac{1}{2}} \hbar \vec{k})}} u^{(r)\dag }(\vec{p}+ {\textstyle \frac{1}{2}} \hbar \vec{k}) \widetilde{\mathcal{V}_{G}}(\vec{k},\vec{p}- {\textstyle \frac{1}{2}} \hbar \vec{k},t)u^{(r^{\prime })}(\vec{p} - {\textstyle \frac{1}{2}} \hbar \vec{k})}   \label{phase} \\
\qquad &=&\frac{c^{2}}{2E(\vec{p})}p^{\mu }\widetilde{h}_{\mu \nu }p^{\nu }\delta
_{rr^{\prime }}-\frac{i\hbar c}{4\gamma }\left[ \vec{k}\times \left( 
\widetilde{\vec{h}}-\widetilde{\overset{\Rightarrow }{h}}\cdot \frac{\vec{p}c%
}{E(\vec{p})}\right) \right] \cdot w^{(r)\dag }\vec{a}w^{(r^{\prime })} 
+\frac{i\hbar }{2m(\gamma +1)}\left[ (\vec{k}\times \vec{p})\frac{%
c^{2}p^{\mu }\widetilde{h}_{\mu \nu }p^{\nu }}{2E^{2}(\vec{p})}\right] \cdot
w^{(r)\dag }\vec{a}w^{(r^{\prime })} \notag
\end{eqnarray}

If we replace now $\vec{p}$ by $\vec{p}+\hbar \vec{k}/2$ in order to
calculate (\ref{amplitude}) this introduces the additional terms:
\begin{equation}
\left[ \frac{\hbar c^{2}\vec{k}\cdot \vec{p}}{4E(\vec{p})}\left( \widetilde{h%
}_{00}-\frac{c^{2}\vec{p}\cdot \widetilde{\overset{\Rightarrow }{h}}\cdot 
\vec{p}}{E^{2}(\vec{p})}\right) -c\frac{\hbar \vec{k}}{2}\cdot \left( 
\widetilde{\vec{h}}-\frac{\widetilde{\overset{\Rightarrow }{h}}\cdot \vec{p}c%
}{E(\vec{p})}\right) \right] \delta _{rr^{\prime }}
\end{equation}

If we introduce the 4-vector $\kappa ^{\mu }$: 
\begin{eqnarray}
\kappa ^{0}c &=&\vec{k}\cdot \vec{v} \\
\vec{\kappa} &=&\vec{k}
\end{eqnarray}
which corresponds to the energy--momentum 4--vector exchanged during the
interaction, these terms can be rewritten: 
\begin{equation}
\frac{c^{2}}{2E(\vec{p})}\left\{ -p^{\mu }\widetilde{h}_{\mu \nu }p^{\nu }%
\frac{\hbar \kappa ^{0}c}{E(\vec{p})}+\frac{\hbar \kappa ^{\mu }}{2}%
\widetilde{h}_{\mu \nu }p^{\nu }\right\} \delta _{rr^{\prime }}
\end{equation}

Our final result is thus, for the scattering amplitude: 
\begin{eqnarray}
\left\langle \chi _{\mathrm{ref}}\right| U^{(1)}(t,t_{0})\left| \chi
(t_{0})\right\rangle & =  & -\frac{i}{\hbar }\sum_{r,r^{\prime
}}\int_{t_{0}}^{t}dt^{\prime }\int (d^{3}p)\int \frac{d^{3}k}{(2\pi )^{3/2}}%
\left\langle \chi _{\mathrm{ref}}\right| 1_{\vec{p}+\hbar \vec{k},r}\rangle 
\left\{ \frac{c^{2}}{2E(\vec{p})}\left( p^{\mu }+\hbar \kappa ^{\mu
}\right) \widetilde{h}_{\mu \nu }p^{\nu }\left( 1-\frac{\hbar \kappa ^{0}c}{%
2E(\vec{p})}\right) \delta _{rr^{\prime }}\right.  \notag \\
&& \qquad\qquad \left. +\frac{i\hbar }{2m(\gamma +1)}\left[ (\vec{k}\times \vec{p})\frac{%
c^{2}p^{\mu }\widetilde{h}_{\mu \nu }p^{\nu }}{2E^{2}(\vec{p})}\right] \cdot
w^{(r)\dag }\vec{a}w^{(r^{\prime })} \right.  \notag \\
&& \qquad\qquad \left. -\frac{i\hbar c}{4\gamma }\left[ \vec{k}\times \left( \widetilde{\vec{h}}-\widetilde{\overset{\Rightarrow }{h}}\cdot \frac{\vec{p}c}{E(\vec{p})}\right)\right] \cdot w^{(r)\dag }\vec{a}w^{(r^{\prime })}\right\} e^{i\left[ E(\vec{p}+\hbar \vec{k})-E(\vec{p})\right] t^{\prime }/\hbar
}\langle 1_{\vec{p},r^{\prime }}\left| \chi _{i}\right\rangle \, .  \label{matrix} 
\end{eqnarray}
To obtain the outgoing spinor, one can replace $\left\langle \chi _{\mathrm{%
ref}}\right| $ by $\left\langle \phi _{0}\right| \theta (x)$ in the previous
expression, which gives this spinor as a sum of outgoing plane--wave spinors: 
\begin{eqnarray}
\chi(x) & = & \chi_{i}(x)-\frac{i}{\hbar }\frac{1}{(2\pi \hbar)^{3/2}} \sum_{r,r^{\prime }} \int_{t_0}^t dt^{\prime }\int d^3 p \int \frac{d^{3}k}{(2\pi )^{3/2}} \sqrt{\frac{mc^{2}}{E(\vec{p})}}u^{(r)}(\vec{p})e^{i\left[ \vec{p}.\vec{x} - E(\vec{p})t/\hbar\right] }  \notag \\
&& \qquad\qquad \left\{ \frac{c^{2}}{2E(\vec{p})}p^{\mu }\widetilde{h}_{\mu \nu }\left(
p^{\nu }-\hbar \kappa ^{\nu }\right) \left( 1+\frac{\hbar \kappa ^{0}c}{2E(\vec{p})}\right) \delta _{rr^{\prime }} +\frac{i\hbar }{2m(\gamma +1)}\left[ (\vec{k}\times \vec{p})\frac{c^{2}p^{\mu }\widetilde{h}_{\mu \nu }p^{\nu }}{2E^{2}(\vec{p})}\right] \cdot
w^{(r)\dag }\vec{a}w^{(r^{\prime })}\right.  \notag \\
&& \qquad\qquad\qquad \left. -\frac{i\hbar c}{4\gamma }\left[ \vec{k}\times \left( \widetilde{%
\vec{h}}-\widetilde{\overset{\Rightarrow }{h}}\cdot \frac{\vec{p}c}{E(\vec{p})}\right) \right] \cdot w^{(r)\dag }\vec{a}w^{(r^{\prime })}\right\} e^{i\left[ E(\vec{p})-E(\vec{p}-\hbar \vec{k})\right] t^{\prime }/\hbar
}\langle 1_{\vec{p}-\hbar \vec{k},r^{\prime }}\left| \chi _{i}\right\rangle
\end{eqnarray}
in which an explicit phase factor is associated with each outgoing plane
wave component and which can also be obtained directly from the first--order
solution of equation (\ref{eqmomentum}).

One can also perform the calculation presented in Appendix D, which gives
this spinor in the form of Dirac matrices multiplying the initial spinor
plane wave components
\begin{eqnarray}
\chi (x) &=&\chi _{i}(x)-\frac{i}{\hbar }\frac{1}{(2\pi \hbar )^{3/2}}%
\sum_{r^{\prime }}\int_{t_{0}}^{t}dt^{\prime }\int (d^{3}p)\int \frac{d^{3}k%
}{(2\pi )^{3/2}} \frac{c^{2}}{2E(\vec{p})}\left\{ \left( p^{\mu }+\hbar \kappa ^{\mu
}\right) \widetilde{h}_{\mu \nu }p^{\nu }\left( 1-\frac{\hbar \kappa ^{0}c}{%
E(\vec{p})}\right) -\frac{i\hbar }{2}\kappa _{\rho }\sigma ^{\rho \nu }%
\widetilde{h}_{\mu \nu }p^{\mu }\right\}  \notag \\
&& \qquad\qquad\qquad\qquad e^{i\vec{k}.\vec{x}}e^{i\left[ E(\vec{p}+\hbar \vec{k})-E(\vec{p})\right]
(t^{\prime }-t)/\hbar } \sqrt{\frac{mc^{2}}{E(\vec{p})}}u^{(r^{\prime })}(\vec{p})e^{i(\vec{p} \cdot \vec{x}-E(\vec{p})t)/\hbar }\langle 1_{\vec{p},r^{\prime }}\left| \chi
_{i}\right\rangle \, . \label{Spinor}
\end{eqnarray}
In this formula, the integral over $\vec{p}$ can be calculated by assuming
that the initial wave packet has a very narrow width in momentum space
around a central value $\vec{p}_{0}$. The initial wave packet $\chi _{i}(x)$
can then be factorized. If this approximation is not sufficient an expansion
of the wave packet around $\vec{p}_0$ can be used \cite{icols99}. The $%
\vec{k}$ integral can be performed by turning each term linear in $\vec{k}$
into a spatial derivative and the result of Appendix C is recovered for the
spinor in configuration space. Finally the time integral can be worked out
in many cases and expresses energy conservation \cite{icols99}.

For the scattering amplitude, the comparison of equation (\ref{matrix}) with
equation (\ref{PSCS}) shows new terms directly related to the momentum
exchange: a generalized Thomas precession and a generalized spin--gravitation
interaction. To illustrate how this phase shift calculation is done from
equation (\ref{matrix}) we shall rewrite this equation without the terms
that obviously do not contribute to the phase and use expression (\ref
{energy}) for the energy difference, in which we neglect the recoil shift $\delta $
\begin{eqnarray}
\lefteqn{\delta \varphi = -\frac{1}{\hbar}\sum_{r,r^{\prime
}}\int_{t_{0}}^{t}dt^{\prime }\int (d^{3}p)\alpha_{\mathrm{ref}}^{\ast }(%
\vec{p})\alpha _{i}(\vec{p})\beta_{r,\mathrm{ref}}^{\ast }\beta_{r^{\prime
},i} \Biggl\{ \frac{c^{2}}{2E(\vec{p})}p^{\mu} h_{\mu\nu}(\vec{x}_0 + \vec{v} t^\prime, t^\prime)p^{\nu }\delta _{rr^{\prime }}}  \nonumber \\
&& \qquad\qquad\qquad\qquad\qquad\qquad +\frac{\hbar}{2m(\gamma +1)}\left[ \frac{c^2 p^{\mu }\vec{\nabla} h_{\mu \nu }(\vec{x}_0 + \vec{v} t^\prime, t^\prime)p^{\nu }}{2E^{2}(\vec{p})}\times \vec{p}\right] \cdot w^{(r)\dag }\vec{a}w^{(r^{\prime
})} \label{calcul} \\
& & \qquad\qquad\qquad\qquad\qquad\qquad -\frac{\hbar c}{4\gamma }\left[ \vec{\nabla}\times \left( \vec{h}(%
\vec{x}_{0}+\vec{v}t^{\prime },t^{\prime })-\overset{\Rightarrow }{h}(\vec{x}%
_{0}+\vec{v}t^{\prime },t^{\prime })\cdot \frac{\vec{p}c}{E(\vec{p})}\right) %
\right] \cdot w^{(r)\dag }\vec{a}w^{(r^{\prime })}\Biggr\} \, ,  \notag 
\end{eqnarray}
where we also made explicit the centers of the wave packets and their
polarization: 
\begin{equation}
\langle 1_{\vec{p},r}\left| \chi_{i}\right\rangle = e^{-i\vec{p}.\vec{x}_{0}/\hbar }\alpha _{i}(\vec{p})\beta _{r,i},\left\langle \chi _{\mathrm{ref}}\right| 1_{\vec{p}+\hbar \vec{k},r}\rangle \simeq e^{i\left( \vec{p}+\hbar 
\vec{k}\right) \cdot \vec{x}_{0}/\hbar }\alpha_{\mathrm{ref}}^{\ast }(\vec{p})\beta _{r,\mathrm{ref}}^{\ast }
\end{equation}
(the assumption that the wave packet is broad enough for $\alpha _{\mathrm{ref}}^{\ast }(\vec{p}+\hbar \vec{k})\simeq \alpha_{\mathrm{ref}}^{\ast
}(\vec{p})$ has been made).

If we assume, for simplicity, that the reference wave packet is identical to
the unperturbed wave packet: $\alpha _{_{\mathrm{ref}}}(\vec{p})\beta _{r,%
\mathrm{ref}}\equiv \alpha _{i}(\vec{p})\beta _{r,i}$ and that they are very
narrow in momentum space around a central value $\vec{p}$, the phase
simplifies to: 
\begin{eqnarray}
\delta \varphi &=&\;-\frac{1}{\hbar }\int_{t_{0}}^{t}dt^{\prime }\left\{ 
\frac{c^{2}}{2E(\vec{p})}p^{\mu }h_{\mu \nu }(\vec{x}_{0}+\vec{v}t^{\prime
},t^{\prime })p^{\nu } +\frac{\gamma}{m(\gamma +1)}\left[ \frac{c^{2}p^{\mu }\vec{\nabla}h_{\mu
\nu }(\vec{x}_{0}+\vec{v}t^{\prime },t^{\prime })p^{\nu }}{2E^{2}(\vec{p})}%
\times \vec{p}\right] \cdot \overline{\vec{s}}\right.  \notag \\
&& \qquad\qquad\qquad\qquad\qquad\qquad\qquad\qquad \left. -\frac{c}{2}\left[ \vec{\nabla}\times \left( \vec{h}(\vec{x}%
_{0}+\vec{v}t^{\prime },t^{\prime })-\overset{\Rightarrow }{h}(\vec{x}_{0}+%
\vec{v}t^{\prime },t^{\prime })\cdot \frac{\vec{p}c}{E(\vec{p})}\right) %
\right] \cdot \overline{\vec{s}}\right\}\, ,  \label{phasefin}
\end{eqnarray}

\endlandscape

where $\overline{\vec{s}}$ is the mean spin vector\footnote{%
More generally, if we use $a_{i}(\vec{p},r)$ instead of $\alpha _{i}(\vec{p})\beta _{r,i}$, the mean spin vector should be written: 
\begin{equation}
\overline{\vec{s}}=\left\langle \chi _{i}\right| \frac{\hbar }{2}\int
(d^{3}x\,)\theta ^{\dagger }(x)\vec{\Sigma}\theta (x)\left| \chi
_{i}\right\rangle =\sum_{r,r^{\prime }}\int (d^{3}p)a_{i}^{\ast }(\vec{p},r)a_{i}(\vec{p},r^{\prime })\hbar w^{(r)\dag }\vec{a}w^{(r^{\prime
})}/2\gamma
\end{equation}
}
\begin{equation}
\overline{\vec{s}}=\sum_{r,r^{\prime }}\beta _{r,i}^{\ast }\beta _{r^{\prime
},i}\hbar w^{(r)\dag }\vec{a}w^{(r^{\prime })}/2\gamma  \label{B_spin}
\end{equation}

In fact the phase calculation is usually more involved since the previous
formula applies only to the case of straight unperturbed trajectories. In
practice however, one cannot always ignore the fact that, when calculating
the phase to first--order for a given term of the Hamiltonian, the motion of
the particles is affected by other terms. One example is the calculation of
the gravitational shift within the atom beam splitters, in which one cannot
ignore the important effects of the diffracting electromagnetic field on the
trajectories of the particles \cite{Borde6,Lam1,Lam2,Lam3}. Gravitational
phase shifts have to be calculated along these trajectories. Another example
is the gravity field itself, which, on earth, gives parabolic trajectories
for atoms. The phase shift for the other terms in equation (\ref{phasefin})
has to be calculated along these parabolas. A convenient way to achieve
these calculations is to replace $\vec{x}_{0}+\vec{v}t^{\prime }$ and $\vec{v%
}$ in equation (\ref{phasefin}) by the classical trajectory $\{\vec{x}%
(t^{\prime }),\vec{v}(t^{\prime })\}$ obtained in the $ABCD$ formalism
developed in references \cite{icols99,Borde7}.

Expression (\ref{phasefin}) displays all the terms which may lead to a
gravitational phase shift in a matter--wave interferometer. They are
summarized in Table 1 where one finds successively:

\begin{itemize}
\item  the terms involving $h_{00}$ lead to the gravitational shift ($%
h_{00}=-2$ $\vec{g}\cdot \vec{r}/c^{2})$, to shifts involving higher
derivatives of the gravitational potential and to the analog of the Thomas
precession \index{Thomas precession} (spin--orbit coupling corrected by the Thomas factor).

\item  the terms which involve $\vec{h}=\{h^{0k}\},$ give the \index{Sagnac effect} \index{Sagnac effect} Sagnac effect
in a rotating frame ($\vec{h}=\vec{\Omega}\times \mathbf{\vec{r}}/c$), the
spin--rotation coupling \index{spin--rotation coupling} and a relativistic correction (analogous to the
Thomas term for $h_{00}$). They describe also the Lense--Thirring effects \index{Lense--Thirring effect} 
coming from inertial frame--dragging \index{frame--dragging} by a massive rotating body, which is a
source for $\vec{h}$.

\item  the other terms, which involve the \ tensor $\overset{\Rightarrow }{h}%
=\{h^{ij}\}$ describe genuine General Relativity effects such as the effect
of gravitational waves and de \index{de Sitter effect|see{geodetic precession}} Sitter geodetic precession \index{geodetic precession} (which also
includes the Thomas term for $h_{00}$).
\end{itemize}

\begin{sidewaystable}
\centering
\renewcommand{\arraystretch}{1.3}
\begin{tabular}{|c|c|c|}
\hline
Corresponding energy term $V$ & $h_{\mu \nu }$ & Name of the effect \\ 
\hline\hline
& Newtonian potential: $h_{00}=2U/c^{2} = - 2 \vec{g} \cdot \vec{x}/c^{2}$ & 
Gravitational red shift  \\ \cline{2-3}
& or acceleration field $h_{00}=2\vec{a} \cdot \vec{x}/c^{2}$ & Acceleration
shift \\ \cline{2-3}
$\frac{1}{2} E h_{00}$ & Gravity gradient $\vec{g}(z) \cdot \vec{x}=-\left( g-g^{\prime
}z/2\right) z$ &  \\ \cline{2-3}
& or curvature $R_{0i0j}x^{i}x^{j}$ &  \\ \cline{2-3}
& Fermi gauge: $h_{00}^{F}=\ddot{h}_{+}(t-z/c).(x^{2}-y^{2})/2  +\ddot{h}_{\times}(t-z/c) x y $ & gravitational waves \\ \hline
$\displaystyle \frac{\gamma}{2m(\gamma +1)}\left(\vec{\nabla} h_{00} \times \vec{p}\right) \cdot \overline{\vec{s}}$ & $h_{00}=2U/c^{2}$ gives $\displaystyle V = \frac{1}{mc^{2}}\frac{\gamma}{\gamma +1}\left[ \vec{\nabla}U \times \vec{p}\right] \cdot \overline{\vec{s}}$ & \ \ Thomas precession \\ [2ex] \hline\hline
$- c \vec{p} \cdot \vec{h}$  & Rotating frame: $\vec{h} = \vec{\Omega}\times \vec{x}/c$ gives $V=-\vec{\Omega} \cdot \vec{L}$ & Sagnac effect \\ 
\cline{2-3}
& $h_{0i}$ given by the Lense--Thirring metric & Lense--Thirring (orbital) \\ 
\hline
$-(c/2)\left[ \vec{\nabla}\times \vec{h}\right] \cdot \overline{\vec{s}}$ & $\vec{h}=\vec{\Omega}\times \vec{x}/c$ gives $V=-\vec{\Omega}\cdot \overline{\vec{s}}$
& Spin-rotation interaction \\ \cline{2-3}
& $h_{0i}$ given by the Lense-Thirring metric & Lense--Thirring (spin) \\ 
\hline
$\displaystyle -\frac{\gamma }{m(\gamma +1)}\left[ \vec{\nabla}\left( c\vec{p}\cdot \vec{h}/E\right) \times \vec{p}\right] \cdot \overline{\vec{s}}$ &  & $\sim $ Thomas for rotation \\ [2ex] \hline\hline
& Schwarzschild metric in isotropic coordinates: &  \\ [-1ex]
& $h_{00}=h_{11}=h_{22}=h_{33}=2U/c^{2}$ gives &  \\ [-1ex]
$c^{2}\vec{p}.\overset{\Longrightarrow }{h}.\vec{p}/2E$ &  $V=$ $p^{2}U/E$
in addition to $EU/c^{2}$ from $h_{00}$ &  \\ \cline{2-3}
& Einstein gauge: $h_{11}=-h_{22}=h_{+}(t-z/c),$ & Effect of gravitational 
\\ [-1ex]
& $h_{12}=h_{21}=h_{\times }(t-z/c)$ & waves \\ \hline
& Schwarzschild metric: $U=-GM/r$ &  \\ [-1ex]
& $h_{00}=h_{11}=h_{22}=h_{33}=2U/c^{2}$  & de Sitter  \\ [-1ex]
$(c/2)\left[ \vec{\nabla}\times \left( \overset{\Longrightarrow }{h}\cdot \vec{p} c/E\right) \right] \cdot \overline{\vec{s}}$ & gives $\displaystyle V=\frac{1}{mc^{2}}\frac{1}{\gamma }\left[ \vec{\nabla}U\times \vec{p}\right] \cdot \overline{\vec{s}}$ \ in addition & or geodetic precession \\ [1ex]
&  to $\displaystyle V=\frac{1}{mc^{2}}\frac{\gamma }{\gamma +1}\left[ \vec{\nabla}U\times 
\vec{p}\right] \cdot \overline{\vec{s}}$ \ from $h_{00}$ &  \\ \cline{2-3}
& Einstein gauge: &  Interaction of the spin \\ [-1ex]
&  $h_{ij}$ & with gravitational waves \\ \hline
$\displaystyle \frac{\gamma }{2m(\gamma +1)}\left[ \vec{\nabla}\left( c^{2}\vec{p} \cdot \overset{\Longrightarrow }{h} \cdot \vec{p}/E^{2}\right) \times \vec{p}\right] \cdot \overline{\vec{s}}$ &  & $\sim $ Thomas for gravitation \\ [2ex] \hline
\end{tabular}
\renewcommand{\arraystretch}{1}
\caption{ Classification of the various energy terms entering the expression of the phase shift. The factor $\gamma$ is the time dilation factor and should not be confused with the PPN parameter ${\gamma}_{PPN}$ which can also be introduced (for de Sitter precession it gives the familiar factor $({\gamma}_{PPN} +1/2)$).}
\end{sidewaystable}

Our expressions are valid for spins 0 and 1/2 and may be conjectured \ to be
valid for arbitrary spin if $\vec{\sigma}/2$ is replaced by the
corresponding spin operator $\vec{S}$.

The reader will find calculations of the phases corresponding to these
various terms in references \cite
{Borde2,Borde16,Borde6,Peters,Aud1,Aud2,Lam4,Wolf}. In these calculations,
one should never forget that the external field $h_{\mu \nu }$ acts not only
on the atoms but also on other components of the experiments, such as
mirrors and laser beams and that, depending on the chosen gauge, additional
contributions may enter in the final expression of the phase which should,
of course, be gauge independent.

In present experiments on the earth gravity measurements, the relative
sensitivity is $\delta g/g\simeq 3.10^{-9}$ after 60 seconds and the
absolute accuracy $5.10^{-9}$\cite{Borde15,Peters}. For rotations, the best
sensitivity achieved up to now is $6.10^{-10}$rad.s$^{-1}$Hz$^{-1/2}$\cite
{Landragin} but these numbers are expected to improve rapidly in the near
future, especially in space experiments, in which general relativistic
effects should become detectable. An accurate measurement of the effect of
gravitation and inertia on antimatter also appears as a possibility already
discussed in reference \cite{Phillips} with a transmission--grating
interferometer, although we believe that an antiatom interferometer using
laser beams for the antihydrogen beam splitters (Ramsey--Bord\'{e}
interferometers) would be better suited for such an experiment. The
formalism introduced in this paper to deal with antiatoms should be useful
to discuss such experiments, especially when coherent beams of antihydrogen
will be produced either by Bose--Einstein condensation and/or by stimulated
bosonic amplification.

\subsection{Analogy with the electromagnetic interaction}

The formulas that we have derived for the transition amplitude and for the
ougoing spinor strongly suggest analogies with the electromagnetic field
case. To emphasize these analogies, let us introduce the following
pseudo--potential 4--vector \footnote{More rigorously, one should introduce \cite{Borde00} 
\begin{equation}
\widetilde{A}^{\mu }=\frac{1}{2}\widetilde{h}^{\mu \nu }(p_{\nu }+\hbar
\kappa _{\nu }/2) \, ,
\end{equation}
which stems directly from a compact form of the interaction Hamiltonian 
\begin{eqnarray}
\mathcal{V}_{G} &=&\frac{c}{4}\alpha ^{\mu }h_{\mu \nu }p^{\nu }+h.c.=\frac{c%
}{4}\left\{ \alpha ^{\mu }h_{\mu \nu },p^{\nu }\right\} _{+} \\
\text{with }p^{0} &=&-\alpha ^{j}p_{j}+\gamma ^{0}mc\text{ and }p_{j}=i\hbar
\partial _{j}
\end{eqnarray}}
\begin{equation}
\widetilde{A}^{\mu }=\frac{1}{2}\widetilde{h}^{\mu \nu }p_{\nu } \, ,
\end{equation}
then the Linet--Tourrenc term, which appears also in the generalized Thomas
precession is simply 
\begin{equation}
\frac{c^{2}}{2E(\vec{p})}p_{\mu }\widetilde{h}^{\mu \nu }p_{\nu }=\frac{1}{%
\gamma }u_{\mu }\widetilde{A}^{\mu } \, ,
\end{equation}
where $u_{\mu }$ is the 4--velocity $p_{\mu }/m$.

The corresponding field 
\begin{equation}
\widetilde{\Phi }_{\mu \nu }=-i\left( \kappa _{\mu }\widetilde{A}_{\nu
}-\kappa _{\nu }\widetilde{A}_{\mu }\right) =(\widetilde{\mbox{\boldmath$E$}}%
/c,\widetilde{\mbox{\boldmath$B$}})
\end{equation}
appears in the outgoing spinor (\ref{Spinor}) through
\begin{equation*}
-\frac{i\hbar }{2}\kappa _{\rho }\sigma ^{\rho \nu }\widetilde{h}_{\mu \nu
}p^{\mu }=\frac{\hbar }{2}\sigma ^{\rho \nu }\widetilde{\Phi }_{\rho \nu }
\end{equation*}
and in the generalized spin-gravitation interaction term 
\begin{equation}
-\frac{i\hbar c}{4\gamma }\left[ \vec{k}\times \left( \widetilde{\vec{h}}-%
\widetilde{\overset{\Rightarrow }{h}}\cdot \frac{\vec{p}c}{E(\vec{p})}%
\right) \right] \cdot w^{(r)\dag }\vec{a}w^{(r^{\prime })}=-\frac{\hbar c^{2}%
}{2\gamma E(\vec{p})}w^{(r)\dag }\vec{a}w^{(r^{\prime })}\cdot \widetilde{%
\vec{B}} \, .
\end{equation}
This new correspondence between the gravitational interaction and the
electromagnetic interaction generalizes the so--called gravitoelectric and
gravitomagnetic interactions introduced by de Witt \cite{deWitt} and Papini 
\cite{Papini}.

%\appendix

\section*{Appendix A: Dirac equation in curved space--time}\index{Dirac equation}

For a given space--time manifold, together with its metric tensor $g_{\mu
\nu }$, the Lagrangian density of the Dirac field reads \cite{W} 
\begin{equation}
\mathcal{L}=\frac{\hbar c}{2}\sqrt{-g}\,\overline{\Psi }\left[ i\gamma
^{\alpha }\overrightarrow{\mathcal{D}}_{\alpha }-mc/\hbar \right] \Psi +%
\frac{\hbar c}{2}\sqrt{-g}\,\overline{\Psi }\left[ -i\overleftarrow{\mathcal{%
D}}_{\alpha }\gamma ^{\alpha }-mc/\hbar \right] \Psi ,  \label{LAG}
\end{equation}
with 
\begin{equation}
\overrightarrow{\mathcal{D}}_{\alpha }=e_{\widehat{\alpha }}\,^{\mu }\left[ 
\overrightarrow{\partial }_{\mu }-\frac{i}{4}e_{\widehat{\beta }}^{\nu
}\nabla _{\mu }e_{\widehat{\gamma }\nu }\sigma ^{\beta \gamma }\right]
\,,\quad \overleftarrow{\mathcal{D}}_{\alpha }=\left[ \overleftarrow{%
\partial }_{\mu }+\frac{i}{4}e_{\widehat{\beta }}^{\nu }\nabla _{\mu }e_{%
\widehat{\gamma }\nu }\sigma ^{\beta \gamma }\right] e_{\widehat{\alpha }%
}\,^{\mu }\,.
\end{equation}
In these formulas the $e_{\widehat{\alpha }}$'s are four-vector fields
constituting an orthonormal tetrad (tetrad indices have a hat), that is
satisfying the relations 
\begin{equation}
g_{\mu \nu }\,e_{\widehat{\alpha }}\,^{\mu }e_{\widehat{\beta }}\,^{\nu
}=\eta _{\alpha \beta },  \label{g}
\end{equation}
with $(\eta _{\alpha \beta })=diag(+1,-1,-1,-1)$. A change of tetrad is
possible according to the formula 
\begin{equation}
e_{\widehat{\alpha }}^{\prime }\,^{\mu }=\Lambda ^{\beta }\,_{\alpha }e_{%
\widehat{\beta }}\,^{\mu },  \label{e'}
\end{equation}
where $\Lambda $ is a matrix of the Lorentz group defined at each point of
the spacetime manifold. In such a change the Lagrangian density (\ref{LAG})
is left invariant if the Dirac field is correspondingly transformed
according to the law 
\begin{equation}
\Psi ^{\prime }=S(\Lambda )^{-1}\Psi ,  \label{Psi}
\end{equation}
the matrix $S(\Lambda )$ being the usual transformation matrix under a
Lorentz transformation of a Dirac spinor \cite{BD}. For an infinitesimal
transformation 
\begin{equation}
\Lambda _{\alpha \beta }=\eta _{\alpha \beta }+\varepsilon _{\alpha \beta },%
\hspace{2cm}\varepsilon _{\alpha \beta }=-\varepsilon _{\beta \alpha },
\label{B_epsilon}
\end{equation}
one has 
\begin{equation}
S(\Lambda )^{-1}=I+\frac{i}{4}\varepsilon _{\alpha \beta }\sigma ^{\alpha
\beta }.  \label{S}
\end{equation}
The field equations read 
\begin{equation}
\left[ i\gamma ^{\alpha }\overrightarrow{\mathcal{D}}_{\alpha }-mc/\hbar %
\right] \Psi =0\,,\qquad \overline{\Psi }\left[ -i\overleftarrow{\mathcal{D}}%
_{\alpha }\gamma ^{\alpha }-mc/\hbar \right] =0.
\end{equation}
With the invariance of $\mathcal{L}$ under the phase transformations $\Psi
\rightarrow \Psi e^{-i\alpha },\overline{\Psi }\rightarrow \overline{\Psi }%
e^{i\alpha },$ is associated the current density 
\begin{equation}
j^{\mu }=\sqrt{-g}\,J^{\mu }\,,\qquad J^{\mu }=ce_{\widehat{\alpha }}^{\mu
}\,\overline{\Psi }\gamma ^{\alpha }\Psi ,  \label{J}
\end{equation}
where $J^{\mu }$ is a four-vector invariant under a change of tetrad, and
one has the conservation relation in either one of the two equivalent forms 
\begin{equation}
\partial _{\mu }j^{\mu }=0\,,\qquad \nabla _{\mu }J^{\mu }=0\,.
\end{equation}

\section*{Appendix B: Weak-Field approximation}

It is now assumed that space-time admits a coordinate system $(x^{\mu })$ in
which the metric tensor takes the form 
\begin{equation}
g_{\mu \nu }=\eta _{\mu \nu }+h_{\mu \nu },\hspace{2cm}|h_{\mu \nu }| \ll 1.
\label{h}
\end{equation}
According to that hypothesis, the $h_{\mu \nu }$'s will be considered as
first-order quantities, and the subsequent calculations will be valid at
this order. To determine the corresponding form of the Lagrangian, it
suffices to construct the tetrads associated with (\ref{h}). By putting $e_{%
\widehat{\alpha }\mu }=\eta _{\alpha \mu }+f_{\alpha \mu },$ where $%
f_{\alpha \mu }$ is of the first order, one obtains from (\ref{g}) and (\ref
{h}), 
\begin{equation}
f_{\alpha \beta }+f_{\beta \alpha }=h_{\alpha \beta }.
\end{equation}
The general solution of these equations is of the form 
\begin{equation}
f_{\alpha\beta} = {\textstyle \frac{1}{2}} h_{\alpha\beta} + \epsilon_{\alpha\beta}\, ,
\end{equation}
in which the $\epsilon _{\alpha \beta }$'s are first-order quantities only
restricted by the antisymmetry condition $\epsilon _{\alpha \beta
}=-\epsilon _{\beta \alpha }$. One then has 
\begin{equation}
e_{\widehat{\alpha }\mu }=\eta_{\alpha\mu} + {\textstyle \frac{1}{2}} h_{\alpha \mu}+\epsilon _{\alpha \mu }.  \label{e}
\end{equation}
Introducing the free Dirac Lagrangian (\ref{lag0}) and the associated
stress-energy tensor (\ref{T}), the Lagrangian density (\ref{LAG})
calculated at the first order and corresponding to (\ref{e}) reads\footnote{%
The indices of $h_{\mu \nu }$ and $\epsilon _{\mu \nu }$ are raised with the
help of $\eta ^{\mu \nu }$.} 
\begin{equation}
\mathcal{L}=\mathcal{L}_0 - {\textstyle \frac{1}{2}} h^{\mu \nu }T_{\mu \nu} + \frac{i\hbar c}{2}\epsilon ^{\mu \nu }\overline{\Psi }(\gamma _{\mu }\overrightarrow{%
\partial }_{\nu }-\overleftarrow{\partial }_{\nu }\gamma _{\mu })\Psi -\frac{%
\hbar c}{8}\partial _{\mu }\epsilon _{\nu \rho }\overline{\Psi }(\gamma
^{\mu }\sigma ^{\nu \rho }+\sigma ^{\nu \rho }\gamma ^{\mu })\Psi .
\end{equation}
The corresponding field equations are the following 
\begin{eqnarray}
0 & = & \lbrack (1+ {\textstyle \frac{1}{2}} h)(i\gamma ^{\mu }\overrightarrow{\partial }_{\mu }-mc/\hbar )-%
\frac{i}{2}h^{\mu \nu }\gamma _{\mu }\overrightarrow{\partial }_{\nu }  
\notag \\
& & \qquad + \frac{i}{4}(\partial _{\mu }h-\partial _{\nu }h^{\nu }\,_{\mu })\gamma
^{\mu}+i\epsilon^{\mu\nu} \gamma_\mu \overrightarrow{\partial }_{\nu }- \frac{1}{4}\partial _{\mu }\epsilon _{\nu \rho }\gamma ^{\mu }\sigma ^{\nu
\rho }]\Psi \,,  \label{psi} \\
0 & = & \overline{\Psi }[(-i\gamma ^{\mu }\overleftarrow{\partial }_{\mu }-mc/\hbar
)(1+ {\textstyle \frac{1}{2}} h)+\frac{i}{2}\overleftarrow{\partial }_{\nu }\gamma _{\mu }h^{\mu \nu
} \notag \\
& & \qquad -\frac{i}{4}(\partial _{\mu }h-\partial _{\nu }h^{\nu }{}_{\mu })\gamma
^{\mu }-i\overleftarrow{\partial }_{\nu }\gamma _{\mu }\epsilon ^{\mu \nu }-%
\frac{1}{4}\partial _{\mu }\epsilon _{\nu \rho }\sigma ^{\nu \rho }\gamma
^{\mu }] \,.  \label{psibar}
\end{eqnarray}
The terms depending on $h_{\mu \nu }$ are identical to those appearing in (%
\ref{eqpsi}) and (\ref{eqpsibar}). The same is true for the current density,
which is now 
\begin{equation}
j^{\mu }=c\overline{\Psi }[\gamma^\mu + {\textstyle \frac{1}{2}} h\gamma ^{\mu } - {\textstyle \frac{1}{2}} h^{\mu\nu}\gamma_\nu -\epsilon ^{\mu \nu }\gamma _{\nu }]\Psi .
\end{equation}

Considered independently of the context, the equations (\ref{psi}) and (\ref
{psibar}) are invariant under the transformations of the Poincar\'{e} group
provided that the $h_{\mu \nu }$'s and the $\epsilon _{\mu \nu }$'s are
transformed like the components of a second--rank tensor, and $\Psi $ by the
corresponding transformation law of a spinor. However, the weak--field
character of $h_{\mu \nu }$ is not conserved by any finite Lorentz
transformation except by the rotations. Moreover, the $h_{\mu \nu }$'s being
the basic quantities, the condition that the $\epsilon _{\mu \nu }$'s are of
first order is naturally interpreted by assuming that these latter are some
linear functions of the former, that is one has 
\begin{equation}
\epsilon _{\mu \nu }=\alpha _{\mu \nu \rho \sigma }h^{\rho \sigma }.
\label{alpha}
\end{equation}
Such a relation is not compatible with the general tensor transformation
law, on account of the symmetric or antisymmetric character of $h_{\mu \nu }$
or $\epsilon _{\mu \nu }$, but can be made compatible with the rotations by
a suitable choice of the coefficients $\alpha _{\mu \nu \rho \sigma }$.

The rotational invariance of (\ref{alpha}) is equivalent to the relations 
\begin{equation}
\alpha ^{\mu ^{\prime }\nu ^{\prime }\rho ^{\prime }\sigma ^{\prime
}}=R_{\mu }\hspace*{0.1mm}^{\mu ^{\prime }}R_{\nu }\hspace*{0.1mm}^{\nu
^{\prime }}R_{\rho }\hspace*{0.1mm}^{\rho ^{\prime }}R_{\sigma
}\hspace*{0.1mm}^{\sigma ^{\prime }}\alpha ^{\mu \nu \rho \sigma },
\end{equation}
where $R$ can be any rotation matrix. The corresponding solutions are given
by\footnote{%
As usual, the latin indices can take the values 1,2 or 3, and $\varepsilon
^{ijk}$ is the completely antisymmetric symbol.} 
\begin{equation}
\alpha ^{0i0j}={\textstyle \frac{1}{2}} A\delta ^{ij}\,,\quad \alpha ^{0ijk}=0\,,\quad
\alpha ^{jk0i}={\textstyle \frac{1}{2}} B\varepsilon ^{ijk}\,,\quad \alpha ^{ijkl}=0\,,
\end{equation}
where $A$ and $B$ are some arbitrary parameters. If, in addition, the
invariance under parity is postulated, one has to take $B=0$, giving 
\begin{equation}
\epsilon _{0i}=-\epsilon _{i0}=-Ah_{0i}\,,\quad \epsilon _{ij}=0\,,
\end{equation}
then, instead of (\ref{e}), 
\begin{equation}
e_{\widehat{0}0}=1+{\textstyle \frac{1}{2}} h_{00}\,,\quad e_{\widehat{0}i}=({\textstyle \frac{1}{2}} -A)h_{0i}\,,\quad e_{\widehat{i}0}=({\textstyle \frac{1}{2}} +A)h_{0i}\,,\quad e_{\widehat{%
i}j}=\eta _{ij}+{\textstyle \frac{1}{2}} h_{ij}\,.
\end{equation}
In particular the choice $A=0$ corresponds to the tetrad 
\begin{equation}
e_{\widehat{\alpha }\mu }=\eta _{\alpha \mu } + {\textstyle \frac{1}{2}} h_{\alpha \mu }\,,
\label{standard}
\end{equation}
which we call the \emph{standard tetrad} associated with $h_{\mu \nu }$. The
corresponding Lagrangian is identical to the Lagrangian (\ref{lag}),
therefore the equations (\ref{eqpsi}) and (\ref{eqpsibar}) are recovered
from (\ref{psi}) and (\ref{psibar}) by letting $\epsilon _{\mu \nu }=0$. We
will continue to designate the corresponding field by $\Psi $.

If we write explicitly the spinorial connection in this weak-field
approximation 
\begin{equation}
-\frac{i}{4}e_{\widehat{\beta }}^{\nu }\nabla _{\mu }e_{\widehat{\gamma }\nu
}\sigma ^{\beta \gamma }=\frac{i}{4}\sigma ^{\lambda \mu }\partial _{\lambda
}h_{\mu \nu }  \label{connexion}
\end{equation}
the equation for $\Psi $ can also be written in a simple form, analogous to
the electromagnetic case, and which will find an interpretation in terms of
gauge fields in the flat space--time approach of the main text 
\begin{equation}
i\gamma ^{\nu }(\overrightarrow{\partial }_{\nu }+\frac{i}{4}\sigma
^{\lambda \mu }\partial _{\lambda }h_{\mu \nu }-\frac{1}{2}h_{\nu }{}^{
\alpha }\overrightarrow{\partial }_{\alpha })\Psi -\frac{mc}{{\hbar }}\Psi
=0.
\end{equation}

Another choice of tetrad can be made in relation with the expression of the
vector current. From (\ref{J}) written in the weak--field case, one has in
general 
\begin{equation}
j^{\mu }/c=(1+ {\textstyle \frac{1}{2}} h)\overline{\Psi }\gamma ^{\mu }\Psi -{\textstyle \frac{1}{2}} h^{\mu \nu }%
\overline{\Psi }\gamma _{\nu }\Psi -\epsilon ^{\mu \nu }\overline{\Psi }%
\gamma _{\nu }\Psi ,
\end{equation}
and then 
\begin{equation}
j^{0}/c=(1+ {\textstyle \frac{1}{2}} h)\overline{\Psi }\gamma ^{0}\Psi -({\textstyle \frac{1}{2}} h^{0\nu
}+\epsilon ^{0\nu })\overline{\Psi }\gamma _{\nu }\Psi .
\end{equation}
This quantity will be proportional to the usual density $\overline{\Psi }%
\gamma ^{0}\Psi $ if one has $\epsilon ^{0i}=-\frac{1}{2}h^{0i}$, which is
obtained with the choice $A=\frac{1}{2}$. Denoting by $\Psi ^{\prime }$ the
corresponding field, one has\footnote{%
Let us note that, in the expression of the conserved charge corresponding to
(\ref{j0}), the volume element $(1+\frac{1}{2}h^{i}%
\hspace*{0.1mm}_{i})d^{3}x $ is, in the first--order approximation, the
3--volume associated with the spatial metric defined from the coordinate
system $(x^{\mu })$\cite{LL}.} 
\begin{equation}
j^{0}=c(1+{\textstyle\frac{1}{2}}h^{i}{}_{i})\overline{\Psi ^{\prime }}%
\gamma ^{0}\Psi ^{\prime }\,.  \label{j0}
\end{equation}
It is then possible to introduce the new field $\Theta $ by \cite{BKT} 
\begin{equation}
\Theta =(1+{\textstyle\frac{1}{2}}h^{i}{}_{i})^{\frac{1}{2}}\Psi ^{\prime
}\simeq (1+{\textstyle\frac{1}{4}}h^{i}{}_{i})\Psi ^{\prime }\,,
\label{theta}
\end{equation}
such that 
\begin{equation}
j^{0}=c\overline{\Theta }\gamma ^{0}\Theta ,
\end{equation}
as in the free--field case. The change of field $\Psi \rightarrow \Psi
^{\prime }$ corresponds to a change of tetrad as defined above, so that,
from (\ref{e}) and (\ref{standard}), one has the formulas (\ref{Psi}), (\ref{B_epsilon}) and (\ref{S}) written with an infinitesimal parameter $\varepsilon $ equal to $-\epsilon $ that is such that $\varepsilon
_{ij}=0,\;\varepsilon _{0i}=-\varepsilon _{i0}=\frac{1}{2}h_{0i}.$ This
gives 
\begin{equation}
\Psi ^{\prime }=(I-{\textstyle\frac{1}{4}}h_{0i}\gamma ^{0}\gamma ^{i})\Psi ,
\end{equation}
then, from (\ref{theta}), 
\begin{equation}
\Theta =(I+{\textstyle\frac{1}{4}}h-{\textstyle\frac{1}{4}}h_{0\mu }\gamma
^{0}\gamma ^{\mu })\Psi .
\end{equation}
The field $\Theta $ is identical to the field introduced in Section 3.2 for
the purpose of defining the quantization and the interaction picture.

\section*{Appendix C: A stationary phase calculation}

The outgoing spinor can be calculated directly in configuration space with
the help of a stationary phase formula. In the particle case, from the
definition 
\begin{equation}
\chi (x)=\langle \phi _{0}|\theta ^{(+)}(x)|\chi (t)\rangle
\end{equation}
and the integral equation 
\begin{equation}
|\chi (t)\rangle =|\chi (t_{0})\rangle -\frac{i}{\hbar c}%
\int_{t_{0}}^{t}(d^{4}x^{\prime })\theta ^{(+)\dag }(x^{\prime })\mathcal{V}%
_{G}(x^{\prime })\theta ^{(+)}(x^{\prime })|\chi (t^{\prime })\rangle
\end{equation}
one can derive, to first order, the expression 
\begin{equation}
\chi (x)=\chi _{i}(x)-\frac{i}{\hbar c}\int_{t_{0}}^{t}(d^{4}x^{\prime
})\langle \phi _{0}|\theta ^{(+)}(x)\overline{\theta ^{(+)}(x^{\prime })}%
|\phi _{0}\rangle \gamma ^{0}\mathcal{V}_{G}(x^{\prime })\langle \phi
_{0}|\theta ^{(+)}(x^{\prime })|\chi _{i}\rangle ,
\end{equation}
which, with the help of standard formulas, can be transformed into 
\begin{align}
\chi(x)& =\chi_{i}(x)-\frac{1}{\hbar c} \int_{t_0}^t d^4 x^\prime S^{(+)}(x-x^{\prime })\gamma^0 \mathcal{V}_G(x^\prime) \chi_{i}(x^{\prime }) \\
& =\chi_{i}(x)-\frac{i}{\hbar c}\sum_{r}\int d^3p \chi_{\vec{p},r}^{(+)}(x)\int_{t_{0}}^{t}(d^{4}x^{\prime })\overline{\chi _{\vec{p},r}^{(+)}(x^{\prime })}\gamma ^{0}\mathcal{V}_{G}(x^{\prime })\chi
_{i}(x^{\prime }).
\end{align}
By introducing (\ref{ampli2}), one finds 
\begin{equation}
\chi (x)=\chi _{i}(x)-\frac{i}{\hbar c}\sum_{r}\int d^3 p \chi _{\vec{p}%
,r}^{(+)}(x)\;i\hbar c\langle \chi _{\vec{p},r}^{(+)}|U^{(1)}(t,t_{0})|\chi
_{i}\rangle .
\end{equation}
In this last expression the matrix element can be submitted to the same
transformations as those leading from (\ref{S1}) to (\ref{S2}), yielding 
\begin{eqnarray}
\chi (x) &=&\chi _{i}(x)-i\sum_{r}\int d^3 p\chi _{\vec{p}%
,r}^{(+)}(x)\int_{t_{0}}^{t} d^4 x^\prime\,\frac{i}{4}h^{\mu \nu
}(x^{\prime })\overline{\chi _{\vec{p},r}^{(+)}(x^{\prime })}\left( \gamma
_{\mu }\overrightarrow{\partial _{\nu }^{\prime }}-\overleftarrow{\partial
_{\nu }^{\prime }}\gamma _{\mu }\right) \chi _{i}(x^{\prime })  \notag \\
&=&\chi _{i}(x)-\frac{i}{4}\int_{t_{0}}^{t} d^4 x^\prime \,h^{\mu \nu
}(x^{\prime })S^{(+)}(x-x^{\prime })\left( \gamma _{\mu }\overrightarrow{%
\partial _{\nu }^{\prime }}-\overleftarrow{\partial _{\nu }^{\prime }}\gamma
_{\mu }\right) \chi _{i}(x^{\prime }).
\end{eqnarray}
By introducing the expression 
\begin{equation}
S^{(+)}(x-x^{\prime })=\frac{ic}{2(2\pi \hbar )^{3}}\int \frac{d^3 p}{E(%
\vec{p})}(\gamma ^{\mu }p_{\mu }+mc)e^{-ip(x-x^{\prime })/\hbar }\,%
\rule[-3mm]{0.1mm}{8mm}_{p_{0}=+E(\vec{p})/c},
\end{equation}
one finally gets for a plane wave with the momentum ${\vec{p}}_{0}$ 
\begin{eqnarray}
\chi (x) &=&\chi _{i}(x)-\frac{ic^{2}}{8\hbar }\int_{t_0}^t dt^\prime \int d^3 x^\prime h^{\mu \nu}(x^\prime)\times  \notag \\
&&\times \frac{1}{(2\pi \hbar )^{3}}\int \frac{d^3 p}{E(\vec{p})}(p+p_{0})_{\nu }(\gamma ^{\rho }p_{\rho }+mc)e^{-ip(x-x^{\prime })/\hbar
}\gamma_\mu\chi_{i}(x^\prime)\,.  \label{CHI}
\end{eqnarray}

\landscape
The elementary stationary phase formula\footnote{%
See, for instance \cite{BOWO}.} applied successively to the integral over $p$%
, then to the integral over $x^{\prime }$, yields the Linet--Tourrenc result 
\cite{LT}, namely : 
\begin{equation}
\chi(x) = \left[1 - \frac{ic^{2}}{2\hbar }\frac{p_{0\mu }p_{0\nu }}{E(\overrightarrow{p_{0}})}\int_{t_0}^t dt^\prime h^{\mu \nu }(\vec{x}- \vec{v}_{0}(t-t^{\prime }),t^\prime)\right] \chi_{i}(x)\, , \qquad \vec{v}_{0}= \frac{\vec{p}_{0}c^{2}}{E(\vec{p}_{0})} \, .
\end{equation}
This expression is the beginning of an asymptotic expansion in powers of $%
\hbar $ of the form 
\begin{equation}
\chi (x)=\left[ 1-\frac{ic^{2}}{2\hbar }\int_{t_{0}}^{t}dt^{\prime }\,F(\vec{%
x}-\vec{v}_{0}(t-t^{\prime }),t^{\prime })\right] \chi _{i}(x),
\end{equation}
in which $F$ is defined by an expansion in integer powers of $\hbar $ whose
first term corresponds to the Linet--Tourrenc formula. This expansion can be
obtained from a generalization of the stationary phase formula given by
H\"{o}rmander \cite{LH}. The application of this general formula to the two
integrals appearing in (\ref{CHI}) yields, after some complicated
calculations, the following expression of the outgoing spinor (to be derived
in the next Appendix by a simpler method) 
\begin{eqnarray}
F &=&\left\{ \frac{p_{\mu }p_{\nu }h^{\mu \nu }}{E(\vec{p})}+\frac{i\hbar}{2E(\vec{p})}\left[ p_{\nu }\partial _{i}h^{\mu \nu }\gamma^i \gamma_\mu +p_{\mu }\partial _{i}h^{\mu i} - p_\mu \frac{v^{i}}{c}\left(\partial_i h^{\mu 0} + \partial_i h^{\mu
\nu}\gamma^0 \gamma_\nu\right) \right.\right.\nonumber\\
& & \qquad\qquad\qquad\qquad \left.\left. + 2 \frac{p_{\mu }p_{\nu }}{E(\vec{p})} v^{i}\partial _{i}h^{\mu \nu } + \frac{c^{2}(t-t^{\prime })}{E(\vec{p})}\left( \delta ^{ij}- \frac{v^{i}}{c}\frac{v^{j}}{c}\right) \partial _{i}\partial _{j}\left(p_{\mu }p_{\nu }h^{\mu \nu }\right) \right] \right\} _{\vec{p}=\vec{p}_{0}}\, .
\end{eqnarray}

\section*{Appendix D: Derivation of the wave function using the momentum
representation}

The method used for the amplitude in the main text can be used to derive the
wave function. The spinorial wave function for one--(anti)particle states is 
\begin{equation}
\chi (x)=\left\langle \phi _{0}\right| \theta (x)\left| \chi (t)\right\rangle \, .
\end{equation}
For particles: 
\begin{eqnarray}
\left\langle \phi _{0}\right| \theta ^{(+)}(x)\left| \chi (t)\right\rangle
& = & \left\langle \phi _{0}\right| \theta ^{(+)}(x)\left| \chi
_{i}\right\rangle -\frac{i}{\hbar }\frac{1}{(2 \pi \hbar)^{3/2}} \sum_{r,r^{\prime }}\int_{t_{0}}^{t}dt^{\prime }\int d^3 p \int \frac{d^{3}k}{(2\pi)^{3/2}} \frac{mc^{2}}{E(\vec{p}+\hbar \vec{k})}u^{(r)}(\vec{p}+\hbar \vec{k}) u^{(r)\dagger }(\vec{p}+\hbar \vec{k}) \notag \\
&& \qquad\qquad\qquad \widetilde{\mathcal{V}}_{G}(\vec{k},\vec{p},t^{\prime })e^{i\vec{k}\cdot\vec{x}} e^{\frac{i}{\hbar}\left[E(\vec{p}+\hbar \vec{k})-E(\vec{p})\right] (t^{\prime }-t)} \sqrt{\frac{mc^{2}}{E(\vec{p})}}u^{(r^{\prime })}(\vec{p})e^{\frac{i}{\hbar} (\vec{p} \cdot \vec{x}-E(\vec{p})t)}\langle 1_{\vec{p},r^{\prime }}| \chi(t^{\prime })\rangle 
\end{eqnarray}
and a similar formula for antiparticles.
To first order we get 
\begin{eqnarray}
\chi ^{(1)}(x) &=&-\frac{i}{\hbar }\frac{1}{(2\pi \hbar )^{3/2}}%
\sum_{r,r^{\prime }}\int_{t_{0}}^{t}dt^{\prime }\int d^3 p \int \frac{%
d^{3}k}{(2\pi )^{3/2}} \frac{mc^{2}}{E(\vec{p}+\hbar \vec{k})}u^{(r)}(\vec{p}+\hbar \vec{k}%
)u^{(r)\dagger }(\vec{p}+\hbar \vec{k})  \notag \\
&& \qquad\qquad \widetilde{\mathcal{V}}_{G}(\vec{k},\vec{p},t^{\prime })e^{i\vec{k} \cdot \vec{x}%
}e^{\frac{i}{\hbar}\left[ E(\vec{p}+\hbar \vec{k})-E(\vec{p})\right] (t^{\prime }-t)} \sqrt{\frac{mc^{2}}{E(\vec{p})}}u^{(r^{\prime })}(\vec{p})e^{\frac{i}{\hbar}(\vec{p} \cdot \vec{x}-E(\vec{p})t)}\langle 1_{\vec{p},r^{\prime }}\left| \chi
_{i}\right\rangle 
\end{eqnarray}
The next idea is to express the propagator which appears in this equation in
order to write the outgoing spinor in the form of Dirac matrices multiplying
the initial spinor plane wave components.

From
\begin{equation}
\sum_{r}u^{(r)}(\vec{p})\overline{u}^{(r)}(\vec{p})=\frac{1}{2mc}\left[
\gamma ^{\mu }p_{\mu }+mc\right]
\end{equation}
one obtains 
\begin{equation}
\sum_{r}\frac{mc^{2}}{E(\vec{p}+\hbar \vec{k})}u^{(r)}(\vec{p}+\hbar \vec{k})u^{(r)\dagger }(\vec{p}+\hbar \vec{k}) = \frac{c}{2E(\vec{p})}\left[ \gamma^\mu \left(p_\mu + \hbar \kappa_\mu\right) + mc\right] \left( 1-\frac{\hbar \kappa^0 c}{E(\vec{p})}\right) \gamma^0 \, ,
\end{equation}
from which, after some algebra, one finds the spinor 
\begin{eqnarray}
\chi ^{(1)}(x) &=&-\frac{i}{\hbar }\frac{1}{(2\pi \hbar )^{3/2}}%
\sum_{r^{\prime }}\int_{t_{0}}^{t}dt^{\prime }\int (d^{3}p)\int \frac{d^{3}k%
}{(2\pi )^{3/2}} \frac{c^{2}}{2E(\vec{p})}\left\{ \left( p^{\mu }+\hbar \kappa ^{\mu
}\right) \widetilde{h}_{\mu \nu }p^{\nu }\left( 1-\frac{\hbar \kappa ^{0}c}{%
E(\vec{p})}\right) -\frac{i\hbar }{2}\kappa _{\rho }\sigma ^{\rho \nu }%
\widetilde{h}_{\mu \nu }p^{\mu }\right\}  \notag \\
&& \qquad\qquad e^{i\vec{k}.\vec{x}}e^{i\left[ E(\vec{p}+\hbar \vec{k})-E(\vec{p})\right]
(t^{\prime }-t)/\hbar} \sqrt{\frac{mc^{2}}{E(\vec{p})}}u^{(r^{\prime })}(\vec{p}) e^{i(\vec{p} \cdot \vec{x}-E(\vec{p})t)/\hbar }\langle 1_{\vec{p},r^{\prime }}\left| \chi
_{i}\right\rangle \, ,
\end{eqnarray}
where the time--dependent exponential can be expanded to any desired order
for recoil shift corrections: 
\begin{eqnarray}
e^{i\left[ E(\vec{p}+\hbar \vec{k})-E(\vec{p})\right] (t^{\prime }-t)/\hbar
} &=&e^{i\vec{k}\cdot \vec{v}(t^{\prime }-t)}\left[ 1+i\hbar \frac{c^2 (t^\prime - t)}{2E(\vec{p})}\left( \vec{k}^{2}-\left( \frac{c\vec{p}%
\cdot \vec{k}}{E(\vec{p})}\right) ^{2}\right) \right]  \simeq e^{i\vec{k}\cdot \vec{v}(t^{\prime }-t)}\left[ 1+i\delta (t^{\prime
}-t)\right] \, .
\end{eqnarray}
Finally one can check that: 
\begin{equation}
\;\left\langle \chi _{\mathrm{ref}}\right| U^{(1)}(t,t_{0})\left| \chi
(t_{0})\right\rangle =\int d^{3}x\,\;\chi _{\mathrm{ref}}^{\dagger }(x)\chi
^{(1)}(x) \, .
\end{equation}
We obtain indeed
\begin{eqnarray}
\int d^{3}x \, \chi_{\mathrm{ref}}^{\dagger }(x)\chi ^{(1)}(x) & = & - \frac{i}{\hbar }\sum_{r,r^\prime}\int_{t_{0}}^t \!\! dt^\prime \!\! \int \!\! d^3 p \!\! \int \!\! \frac{d^3 k}{(2\pi )^{3/2}}e^{\frac{i}{\hbar}\left[ E(\vec{p}+\hbar \vec{k})-E(\vec{p})\right] t^\prime}  \sqrt{\frac{mc^{2}}{E(\vec{p}+\hbar \vec{k})}}u^{(r)\dagger }(\vec{p} +\hbar \vec{k})\ \left\langle \chi _{\mathrm{ref}}\right| 1_{\vec{p}+\hbar \vec{k},r}\rangle  \notag \\
& & \frac{c^{2}}{2E(\vec{p})}\left\{ \left( p^{\mu }+\hbar \kappa^\mu\right) \widetilde{h}_{\mu \nu }p^{\nu }\left( 1-\frac{\hbar \kappa ^{0}c}{E(\vec{p})}\right) -\frac{i\hbar }{2}\kappa_\rho \sigma^{\rho\nu} \widetilde{h}_{\mu \nu }p^{\mu }\right\} \sqrt{\frac{mc^{2}}{E(\vec{p})}}u^{(r^{\prime })}(\vec{p})\langle 1_{\vec{p},r^{\prime }}\left| \chi _{i}\right\rangle \, , 
\end{eqnarray}
with 
\begin{eqnarray}
& & \sqrt{\frac{mc^{2}}{E(\vec{p}+\hbar \vec{k})}}u^{\dagger (r)}(\vec{p}+\hbar \vec{k})  \frac{c^{2}}{2E(\vec{p})}\left\{ \left( p^{\mu }+\hbar \kappa^\mu\right) \widetilde{h}_{\mu \nu }p^{\nu }\left( 1-\frac{\hbar \kappa ^{0}c}{E(\vec{p})}\right) -i\frac{\hbar}{2} \kappa_\rho \sigma^{\rho\nu} \widetilde{h}_{\mu \nu }p^{\mu }\right\} \sqrt{\frac{mc^{2}}{E(\vec{p})}}u^{(r^{\prime })}(\vec{p}) \notag \\
& & \qquad\qquad\qquad = \frac{c^{2}}{2E(\vec{p})}\left( p^{\mu }+\hbar \kappa ^{\mu }\right) 
\widetilde{h}_{\mu \nu }p^{\nu }\left( 1-\frac{\hbar \kappa ^{0}c}{2E(\vec{p})}\right) \delta _{rr^{\prime }} +\frac{i\hbar }{2m(\gamma +1)}\left[ (\vec{k}\times \vec{p})\frac{c^{2}p^{\mu }\widetilde{h}_{\mu \nu }p^{\nu }}{2E^{2}(\vec{p})}\right] \cdot
w^{(r)\dag }\vec{a}w^{(r^{\prime })} \\
& & \qquad\qquad\qquad\qquad\qquad\qquad\qquad   - \frac{i\hbar c}{4\gamma }\left[ \vec{k}\times \left( \widetilde{\vec{h}} - \widetilde{\overset{\Rightarrow }{h}}\cdot \frac{\vec{p}c}{E(\vec{p})}\right) \right] \cdot w^{(r)\dag }\vec{a}w^{(r^{\prime })}\, , \notag 
\end{eqnarray}
from which equation (\ref{matrix}) is recovered.

\endlandscape

\subsubsection{Acknowledgement:}

One of us (Ch.J. B.) would like to thank Dr.\ C.\ L\"{a}mmer\-zahl and Prof.
Ph. Tourrenc for many stimulating discussions and Prof.\ Dr.\ W.\ Ertmer for
his hospitality at the University of Hannover within the
Sonderforschungsbereich 407.


\begin{thebibliography}{99}
\bibitem{Laurent}  P. Laurent {\it et al.}: \textit{Cold atom clocks on earth and in space}, in R. Blatt, J. Eschner, D. Leibfried and F. Schmidt--Kaler (eds.): \textit{Laser Spectroscopy}, Proceedings of the 14th International Conference on Laser Spectroscopy (World Scientific, Singapore 1999), p.~41.

\bibitem{Borde1} P. Berman (ed.): \textit{Atom Interferometry} (Academic
Press 1997).

\bibitem{Borde2}  Ch.J. Bord\'{e}: Atomic interferometry and laser
spectroscopy, in: \textit{Laser Spectroscopy X} (World Scientific, Singapore 1991), p.~239.

\bibitem{Borde4}  U. Sterr {\it et al.}: Atom interferometry based on
separated light fields, in \cite{Borde1} and \textit{Appl. Phys.} \textbf{B 54}, 341 (1992).

\bibitem{Borde14}  F. Riehle, Th. Kisters, A. Witte, J. Helmcke and Ch.J.
Bord\'{e}: Optical Ramsey Spectroscopy in a Rotating Frame: Sagnac
Effect in a Matter-Wave Interferometer, \textit{Phys. Rev. Lett.} \textbf{67},
177-180 (1991).

\bibitem{Borde15}  B.C. Young, M. Kasevich and S. Chu: Precision
atom interferometry with light pulses, in \cite{Borde1} and references
therein.

\bibitem{Borde16}  Ch.J. Bord\'{e}: Atomic interferometry with
internal state labelling, \textit{Phys. Lett.} \textbf{A 140}, 10 (1989).

\bibitem{Borde17}  Ch.J. Bord\'{e} \textit{et al.}: Optical Ramsey fringes
with travelling waves, \textit{Phys. Rev.} \textbf{A 30}, 1836 (1984).

\bibitem{Borde5}  Ch.J. Bord\'{e} \textit{et al.}: Molecular Interferometry
Experiments, \textit{Phys. Lett.} \textbf{A 188}, 187 (1994).

\bibitem{Borde6}  Ch.J. Bord\'{e}: Matter--wave interferometers: a
synthetic approach, in \cite{Borde1}.

\bibitem{Borde18} For an early treatment of a gravitational wave detector
using an atom interferometric gradiometer see: Ch.J. Bord\'{e}, J. Sharma,
Ph. Tourrenc and Th. Damour: Theoretical approaches to laser
spectroscopy in the presence of gravitational fields, \textit{J. Physique Lettres} \textbf{44}, L-983 (1983).

\bibitem{BKT}  Ch.J. Bord\'{e}, A. Karasiewicz and Ph. Tourrenc: General relativistic framework for atomic interferometry, \textit{Int. J. of Mod.
Phys.} \textbf{D 3}, 157 (1994).

\bibitem{19}  F.W. Hehl and Wei--Tou Ni: Inertial effects of a Dirac
particle, \textit{Phys. Rev.} \textbf{D 42}, 2045 (1990).

\bibitem{W}  S. Weinberg: \textit{Gravitation and Cosmology} (John Wiley and
Sons, New York 1972).

\bibitem{G1}  S.N. Gupta: Quantization of Einstein's gravitational
field: linear approximation, \textit{Proc. Phys. Soc.} \textbf{A 65}, 161 (1952).

\bibitem{G2}  S.N. Gupta: Quantization of Einstein's gravitational
field: general treatment, \textit{Proc. Phys. Soc.} \textbf{A 65}, 608 (1952).

\bibitem{BD}  J.D. Bjorken and S.D. Drell: \textit{Relativistic Quantum
Mechanics}, (McGraw--Hill, New York 1964).

\bibitem{BD2}  J.D. Bjorken and S.D. Drell: \textit{Relativistic Quantum
Fields} (McGraw--Hill, New York 1965).

\bibitem{F}  R.P. Feynman, F.B. Morinigo and W.G. Wagner: \textit{Feynman
Lectures on Gravitation,} edited by B. Hatfield (Addison--Wesley, Reading MA
1995).

\bibitem{LL}  L.D. Landau and E.M. Lifschitz: \textit{The Classical Theory
of Fields} (Addison--Wesley, Reading MA 1951).

\bibitem{SSS} S.S. Schweber: \textit{An Introduction to Relativistic
Quantum Field Theory} (Harper and Row, New York 1961).

\bibitem{icols99} Ch.J. Bord\'{e}: Quantum theory of clocks and of
gravitational sensors using atom interferometry, in R. Blatt, J. Eschner, D. Leibfried and F. Schmidt--Kaler (eds.): \textit{Laser Spectroscopy}, Proceedings of the 14th International Conference on Laser Spectroscopy, (World Scientific, Singapore 1999), p.~160.

\bibitem{LT}  B. Linet et P. Tourrenc: Changement de phase dans un
champ de gravitation: possibilit\'{e} de d\'{e}tection interf\'{e}rentielle, \textit{Can. J. Phys.} \textbf{54}, 1129 (1976).

\bibitem{BOWO}  M. Born and E. Wolf: \textit{Principles of Optics}, third
edition (Pergamon Press, Oxford 1964).

\bibitem{LH}  L. H\"{o}rmander: \textit{The Analysis of Linear Partial
Differential Operators}, Vol. 1, (Springer--Verlag, Berlin 1983).

\bibitem{LL2}  V.B. Berestetskii, E.M. Lifschitz and L.P. Pitaevskii: 
\textit{Relativistic Quantum Theory} (Pergamon Press, Oxford 1971).

\bibitem{BW}  V. Bargmann and E.P. Wigner: Group theoretical
discussion of relativistic wave equations, \textit{Proc. Nat. Acad. Sci.} (USA) 
\textbf{34}, 211 (1948).

\bibitem{fusion}  L. de Broglie: \textit{Th\'{e}orie G\'{e}n\'{e}rale des
particules \`{a} spin (M\'{e}thode de fusion)} (Gauthier--Villars, Paris
1943).

\bibitem{Durand}  E. Durand: 16--component theory of the spin--1
particle and its generalization to arbitrary spin, \textit{Phys. Rev.} \textbf{D 11}, 3405 (1975).

\bibitem{Thomas}  L.H. Thomas: The Kinematics of an Electron with an
Axis, \textit{Phil. Mag.} \textbf{3}, 1 (1927).

\bibitem{PL}  J.K. Lubanski: Sur la th\'{e}orie des particules
\'{e}l\'{e}mentaires de spin quelconque.I., \textit{Physica} \textbf{9}, 310 (1942).

\bibitem{JR}  J.M. Jauch and F. Rohrlich: \textit{The Theory of Photons and
Electrons} (Addison--Wesley, Cambridge MA 1955).

\bibitem{Borde00}  Ch.J. Bord\'{e}: \textit{A comparison of electromagnetic
and weak gravitational interactions in matter--wave interferometry}, to be
published.

\bibitem{Bron}  I.N. Bronshtein and K.A. Semendyayev: \textit{Handbook of
Mathematics} (Springer, Berlin 1997), p.~770.

\bibitem{Peters}  A. Peters, K. Y. Chung and S. Chu: A measurement
of gravitational acceleration by dropping atoms, \textit{Nature} \textbf{400}, 849 (1999).

\bibitem{Landragin}  A. Landragin, T. L. Gustavson and M. A. Kasevich: 
Precision atomic gyroscope, in R. Blatt, J. Eschner, D. Leibfried and F. Schmidt--Kaler (eds.): \textit{Laser Spectroscopy}, Proceedings of
the 14th International Conference on Laser Spectroscopy (World Scientific, Singapore 1999), p.~170.

\bibitem{Borde7}  Ch.J. Bord\'{e}: \textit{Propagation of Laser beams and of
atomic systems}, in J. Dalibard {\it et al.} (eds.): {\it Fundamental Systems in Quantum Optics},(Elsevier 1991), p.~287.

\bibitem{Lam1}  Ch.J. Bord\'{e} and C. L\"{a}mmerzahl: Atom beam
interferometry as two--level particle scattering by a periodic potential,
\textit{Ann. Phys.} (Leipzig) \textbf{8}, 83 (1999).

\bibitem{Lam2}  C. L\"{a}mmerzahl and Ch.J. Bord\'{e}: Rabi
oscillations in gravitational fields: exact solution, \textit{Phys. Lett.} 
\textbf{A 203}, 59 (1995).

\bibitem{Lam3}  C. L\"{a}mmerzahl and Ch.J. Bord\'{e}: Atom
interferometry in gravitational fields: influence of gravitation on the beam
splitter, \textit{Gen. Rel. Grav.}, \textbf{31}, 635 (1999).

\bibitem{Aud1}  J. Audretsch and K.--P. Marzlin: Atom interferometry
with arbitrary laser configurations: exact phase shift for potentials
including inertia and gravitation, \textit{ J. Phys. II} (France) \textbf{4}, 2073 (1994).

\bibitem{Aud2}  J. Audretsch and K.--P. Marzlin: Ramsey fringes in
atomic interferometry: measurability of the influence of space--time curvature, \textit{Phys. Rev.} \textbf{A 50}, 2080 (1994).

\bibitem{Lam4} C. L\"{a}mmerzahl: Relativistic treatment of Raman
light--pulse atom beam interferometer with applications in gravity theory,
\textit{J. Phys. II} (France) \textbf{4}, 2089 (1994).

\bibitem{Wolf}  P. Wolf and Ph. Tourrenc: Gravimetry using atom
interferometers: some systematic effects, \textit{Phys. Lett.} \textbf{A 251}, 241 (1999).

\bibitem{deWitt}  B.S. DeWitt: Superconductors and gravitational drag, \textit{Phys. Rev. Lett.} \textbf{16}, 1092 (1966)

\bibitem{Papini}  G. Papini: Particle wave functions in weak
gravitational fields, \textit{Nuovo Cimento} \textbf{52B}, 136 (1967).

\bibitem{Phillips}  T.J. Phillips: Measuring the gravitational
acceleration of antimatter with an antihydrogen interferometer, \textit{Hyperfine Interactions} \textbf{100}, 163 (1996).
\end{thebibliography}
\end{document}